\documentclass[aps,prl,twocolumn,showpacs,superscriptaddress,longbibliography,floatfix]{revtex4-1}
\usepackage{amssymb}
\usepackage{bm}
\usepackage{amsmath}
\usepackage{graphicx}
\usepackage{epstopdf}
\usepackage{subfigure}
\usepackage{natbib}
\usepackage{epsfig}
\usepackage{amsfonts}
\usepackage{mathrsfs}
\usepackage{comment}
\usepackage{xcolor}
\usepackage[dvipsnames]{xcolor}
\usepackage[toc,page,title,titletoc,header]{appendix}
\usepackage[colorlinks,linkcolor=blue,citecolor=blue,anchorcolor=blue]
{hyperref}
\usepackage{dsfont,amsthm,amsbsy}

\newcommand{\br}{{\bf r}}
\newcommand{\bP}{{\bf P}}

\begin{document}

\title{Microscopic Modeling of the Charge-Density-Waves in the Rare-Earth Tritellurides}

\author{Sijia Zhao}
\email{sijiazgl@stanford.edu}
\affiliation{Department of Applied Physics, Stanford University, Stanford, CA 94305, USA}
\affiliation{Stanford Institute for Materials and Energy Sciences, SLAC National Accelerator Laboratory, Menlo Park, CA 94025}
\author{Julian May-Mann}
\affiliation{Department of Physics, Stanford University, Stanford, CA 94305, USA}
\affiliation{Stanford Institute for Materials and Energy Sciences, SLAC National Accelerator Laboratory, Menlo Park, CA 94025}
\author{Ian R. Fisher}
\affiliation{Department of Applied Physics, Stanford University, Stanford, CA 94305, USA}
\affiliation{Stanford Institute for Materials and Energy Sciences, SLAC National Accelerator Laboratory, Menlo Park, CA 94025}
\author{Steven A. Kivelson}
\email{kivelson@stanford.edu}
\affiliation{Department of Physics, Stanford University, Stanford, CA 94305, USA}
\affiliation{Stanford Institute for Materials and Energy Sciences, SLAC National Accelerator Laboratory, Menlo Park, CA 94025}

\date{\today}

\begin{abstract}
Despite being arguably the simplest and best characterized quasi-2D charge-density-wave (CDW) systems, the rare-earth tritellurides ($R$Te$_3$) continue to yield surprising experimental results, including recent evidence suggestive of mirror-symmetry breaking associated with the onset of CDW order. Motivated by this, we consider a 2D electron-phonon model for a single Te square-net plane, which we analyze using mean-field theory. For an appropriate region of parameter space, we find a finite-temperature continuous transition from the normal state to a unidirectional CDW state with an ordering vector matching that observed experimentally. At lower temperatures, we find a second translation-symmetry-breaking transition, similar to what occurs in $R$Te$_3$ compounds with heavier rare-earth elements. In certain parameter regimes, we also find an intervening mirror-symmetry-breaking transition occurring between the two transitions described above. These results reveal an intrinsic susceptibility to mirror-symmetry breaking in the unidirectional CDW phase, which is relevant to understanding recent experiments on the $R$Te$_3$ compounds.
\end{abstract}

\maketitle

\textit{Introduction}---Charge-density-waves (CDWs) are characterized by a particular pattern (commensurate or incommensurate) of
translation symmetry breaking. The rare-earth tritellurides ($R$Te$_3$)\cite{Ru2006,Ru2008,Sinchenko2015Anisotropy,Ru2008thesis,Zong2021,DiMasi1995,Malliakas2005,Laverock2005,Chaudhuri2025,Yusupov2008,Lavagnini2010,Pfuner2010,Moore2016,Trigo2021,Chen2014,Straquadine2022,Singh2024Tetragonality,GalloFrantz2024,Gweon1998,Kim2006,Robertson2006,DiMasi1994,Sacchetti2009,Tomic2009,Walmsley2020,Eiter2013,Siddique2024,Sinchenko2016} are a particularly simple and well characterized family of CDW systems. They are quasi-two-dimensional (2D) and host stripe-like unidirectional CDWs\cite{STM2007} that form at high temperatures (exceeding room temperature in certain cases). From a theoretical perspective, the $R$Te$_3$ compounds appear to be ideal CDW systems that can be well understood within a weak-coupling framework. This perspective is motivated, in part, by angle-resolved photoemission spectroscopy (ARPES) measurements\cite{ARPES2004,ARPES2008}, which show broad bands with well-defined quasi-particle excitations, even above the CDW transition, indicating that a Fermi-liquid description is justified.  Additionally, the $R$Te$_3$s have remarkably well nested Fermi-surfaces 
that allow CDW formation at relatively weak coupling.

Given this simple structure, one would expect few surprises concerning the nature of the CDW order. There are often two distinct CDW phases - a  high temperature unidirectional incommensurate CDW phase  that onsets at $T_{c1}$ and a low temperature  phase with bidirectional order that onsets at $T_{c2} < T_{c1}$\cite{Ru2008,Moore2010,Banerjee2013,Hu2011,Hu2014}. However, recent Raman experiments have shown unexpected evidence that the high-temperature CDW breaks additional mirror symmetries and hosts ferroaxial order\cite{Singh2025,Wang2022,freitas2026revealingnaturechargedensity}. Although coexisting CDW and
mirror-breaking orders are allowed in principle, Landau theory constrains how such a coexistence state can emerge. For an orthorhombic
$R$Te$_3$ crystal, the following three observations are mutually
incompatible: (1) The CDW immediately below $T_{c1}$ breaks mirror symmetry. 
(2) its wavevector is incommensurate and lies along a mirror
plane of the crystal; (3) the transition at  $T_{c1}$ is continuous.

Experimentally, the transition at $T_{c1}$ appears to be continuous, and X-ray measurements have established (2); moreover, no additional
transition has been reported between $T_{c1}$ and $T_{c2}$\cite{Ru2008,Singh2024Tetragonality}. In principle, a single continuous transition is possible with fine-tuning, but it would generically be unstable to splitting. The inconsistency between phenomenological Landau theory and experimental observations calls for a re-examination of the $R$Te$_3$ CDWs at a microscopic level.

Here, we theoretically investigate the CDW phases of the $R$Te$_3$ using a
microscopic electron--phonon model. Over a range of parameters, the leading
instability we find is to a unidirectional incommensurate CDW with an ordering vector $\bm{Q} \approx  (2k_F, 2k_F)$ relative to the square lattice, where $k_F = 2\pi \times 0.36$ is the characteristic Fermi-momentum of the Te bands. The resulting wavevector
agrees well with experiment and reflects a ``hidden'' one-dimensional electronic structure of the square Te nets\cite{PhysRevB.74.125115}.

To determine whether mirror symmetry is broken deep in the CDW phase, we approximate, for numerical convenience, the incommensurate ordering vector by the nearby
commensurate value $\mathbf Q=2\pi(3/8,3/8)$, corresponding to a
period-8 structure, and minimize the adiabatic free energy over the static displacements within the resulting supercell. Immediately below the CDW transition,
the system retains one mirror and one glide symmetry; the latter is functionally equivalent to a second, perpendicular mirror symmetry when compared to experiment. Interestingly, we find that the commensurate CDW possesses an unusually soft mirror-odd sliding mode, making mirror-symmetry-breaking distortions of the CDW profile very low in energy. 

Cooling below the high-temperature CDW transition, we find two
ordering sequences depending on the phonon shear stiffness. For larger stiffness, a single first-order transition leads to a low-temperature
bidirectional CDW with broken glide symmetry. For smaller stiffness, this transition splits into an intermediate continuous glide-breaking transition, and a lower-temperature first-order transition into the bidirectional CDW state. The final low-temperature phase is the same, independent of the shear stiffness. The mirror symmetry that acts perpendicular to the glide symmetry is unbroken in both ordering sequences.

\begin{figure}[t]
\centering
\includegraphics[width=0.9\linewidth]{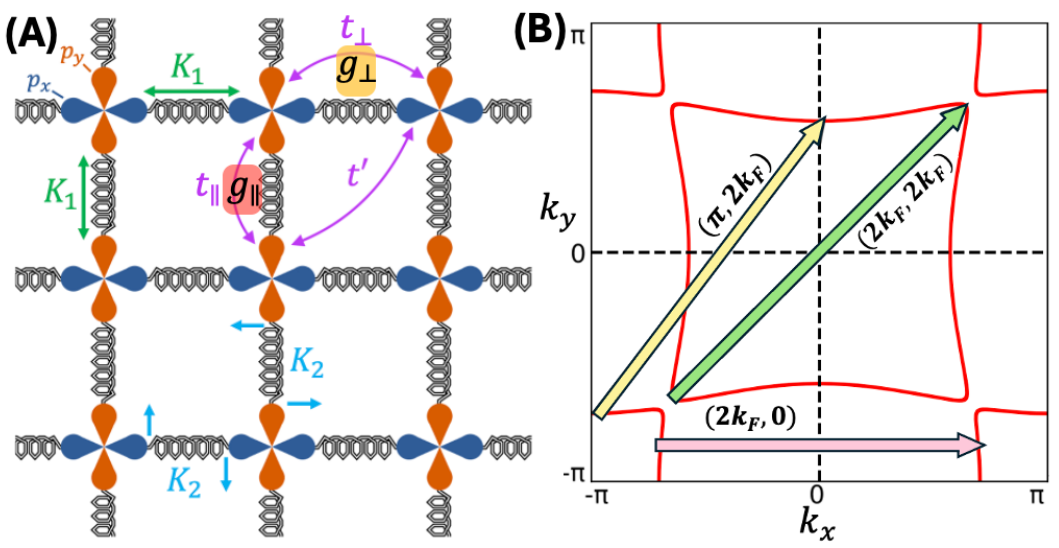}
\caption{\label{fig:model}\textbf{(A)} Schematic of the single-layer Te-square-net model. Here, $t_{\parallel}$, $t_\perp$, and $t'$ denote the three hopping amplitudes,
$g_{\parallel,\perp}$ the corresponding SSH couplings, and
$K_{1,2}$ the longitudinal and transverse stiffnesses. \textbf{(B)} Tight-binding Fermi surface with competing nesting vectors.}
\end{figure}

Our microscopic model reveals that the CDW below $T_{c1}$ is intrinsically susceptible to mirror- and glide-symmetry breaking. This suggests a possible resolution of the apparent conflict between experiments and theoretical expectations: the effect of weak explicit mirror symmetry-breaking perturbations, e.g., strain, might be strongly amplified below $T_{c1}$ while remaining fully consistent with the constraints of Landau theory.

\textit{Model and band structure}---The quasi-two-dimensional orthorhombic $R$Te$_3$ compounds consist of
bilayers of nearly square Te nets separated by insulating rare-earth
tellurium spacers. We focus on a single Te net, where the relevant degrees of freedom are the 5$p_x$ and 5$p_y$ electrons. This is justified by weak inter-net hopping and large $b$-axis resistivity\cite{Sinchenko2015Anisotropy,Ru2006,Ru2008}. It is worth noting that, while $R$Te$_3$ is orthorhombic, each Te net is approximately tetragonal.

We model the CDW as arising from electron-phonon coupling within the square lattice net. Although a CDW mediated by electron-electron interactions is also possible\cite{PhysRevB.110.205103}, ARPES observes well-defined quasiparticle dispersions in good agreement with first-principles band calculations\cite{ARPES2008}, suggesting that electron-electron interactions are not essential and will be neglected here.

The Hamiltonian for a single Te layer is given by
\begin{equation}
    \hat{H} = \hat{H}_e + \hat{H}_{e-ph} + \hat{H}_{ph}
\label{eq:FullModel}
\end{equation}
where the three terms describe the bare electrons, electron--phonon
coupling, and bare phonons. Following previous
electronic-structure studies\cite{kikuchi1998electronic,HY,PhysRevB.110.205103}, we take
\begin{equation}
\begin{split}
\label{Eq1}
&\hat{H}_{e}(\mathbf{k}) = \frac{1}{V}\sum_{\mathbf{k},\sigma = \updownarrow}\begin{pmatrix} \hat{c}^\dagger_{\mathbf{k},p_x, \sigma} & \hat{c}^\dagger_{\mathbf{k},p_y,\sigma}\end{pmatrix} h(\mathbf{k}) \begin{pmatrix} \hat{c}_{\mathbf{k},p_x, \sigma}\\ \hat{c}_{\mathbf{k},p_y,\sigma}\end{pmatrix}\\
&h(\mathbf{k}) = \begin{bmatrix} 2t_{\parallel} \cos k_x \!-\! 2t_{\perp} \cos k_y & 4t' \sin k_x \sin k_y \\ 4t' \sin k_x \sin k_y & \!\!\!\!\! 2t_{\parallel} \cos k_y \!-\! 2t_{\perp} \cos k_x \!\end{bmatrix}
\end{split}
\end{equation}
Here, $\hat{c}^\dagger_{\mathbf k,p_\alpha,\sigma}$ creates an electron with momentum $\mathbf{k}$, spin $\sigma$ and orbital $p_\alpha$ ($\alpha=x,y$). We take $t_\parallel\equiv t=2.0~\mathrm{eV}$, $t_\perp=0.185~t$, and $t'=0.04~t$; these denote the nearest-neighbor $\sigma$-bonding, $\pi$-bonding, and diagonal hopping amplitudes\cite{kikuchi1998electronic}. The model is shown schematically in Fig.~\ref{fig:model}(A). It captures the experimentally relevant mirror symmetries, $M_{x\mp y}$, which send $x\rightarrow \pm y$ (note that the unit cell considered here is rotated by 45$^\circ$ relative to that of $R$Te$_3$). The tetragonal approximation implies that our model has additional $C_4$, $M_x$ and $M_y$ symmetries, which are weakly broken in $R$Te$_3$; 
however, these additional symmetries are unimportant in our analysis.

At the $R$Te$_3$-relevant filling of $n = 2.5$ electrons per Te site, the non-interacting Fermi surface consists of two orthogonal quasi-1D structures due to the highly anisotropic $p_x$ and $p_y$ hoppings. This geometry suggests three possible symmetry-inequivalent CDW wavevectors, $(2k_F, 2k_F)$, $(2k_F,\pi)$, and $(2k_F,0)$, as shown in Fig.~\ref{fig:model}(B). In the strict quasi-1D limit, $t_\perp = t' = 0$, all three nest the Fermi surface and generate CDW instabilities, whereas finite $t_\perp$ and $t'$ select $(2k_F, 2k_F)$ as the leading instability\cite{HY}.

Since the Te electrons couple linearly to in-plane distortions and quadratically to out-of-plane distortions, we retain only the in-plane modes. The resulting SSH-type coupling is
\begin{equation}
\begin{split}
\label{Eq:HamEPh}
\hat{H}_{e-ph} = - \!\!\sum_{\substack{\mathbf{r}, \sigma}} g_{\parallel} &\Big[ (u_{\mathbf{r}+\hat{x},x}-u_{\mathbf{r},x})\hat{c}^\dagger_{\mathbf{r}+\hat{x}, p_x, \sigma} \hat{c}_{\mathbf{r}, p_x,\sigma}   \\ 
&\!\!\!\!\!\!\!\!\!\!\!\!\!\!\!\!\!\!\!\!+ (u_{\mathbf{r}+\hat{y},y}-u_{\mathbf{r},y})\hat{c}^\dagger_{\mathbf{r}+\hat{y}, p_y, \sigma} \hat{c}_{\mathbf{r}, p_y,\sigma} \Big]\\
&\!\!\!\!\!\!\!\!\!\!\!\!\!\!\!\!\!\!\!\!+g_{\perp} \Big[ (u_{\mathbf{r}+\hat{y},y}-u_{\mathbf{r},y}) \hat{c}^\dagger_{\mathbf{r}+\hat{y}, p_x, \sigma} \hat{c}_{\mathbf{r}, p_x,\sigma}   \\ 
&\!\!\!\!\!\!\!\!\!\!\!\!\!\!\!\!\!\!\!\!+(u_{\mathbf{r}+\hat{x},x}-u_{\mathbf{r},x}) \hat{c}^\dagger_{\mathbf{r}+\hat{x}, p_y, \sigma} \hat{c}_{\mathbf{r}, p_y,\sigma} \Big]+ \text{H.c.},\\
\end{split}
\end{equation}
where $\mathbf{r}$ labels the sites of a square lattice with unit vectors $\hat{x}$ and $\hat{y}$. The displacement of the Te atom at site $\bm{r}$ is given by $\bm{u}_{\bm{r}}$. There are two electron-phonon couplings, $g_{\parallel}$ and $g_{\perp}$, which reflect the coupling of $p_{x(y)}$ orbitals to $x(y)$ and $y(x)$ atomic displacements, respectively. We measure the displacements in units of the nearest-neighbor Te--Te spacing $a$, which is set to unity ($a\approx 3.1~\text{\AA}$ in $R$Te$_3$).

Working in the extreme adiabatic limit, where quantum fluctuations of the phonons are neglected, the bare Hamiltonian for the SSH phonon is given by 
\begin{equation}
\begin{split}
\label{Eq:HamPh}
\hat{H}_{ph} = \sum_{\mathbf{r}} &\frac{K_1}{2} \left[ (u_{\mathbf{r}+\hat{x},x}-u_{\mathbf{r},x})^2 + (u_{\mathbf{r}+\hat{y},y}-u_{\mathbf{r},y})^2\right]\\   + &\frac{K_2}{2} \left[ (u_{\mathbf{r}+\hat{x},y}-u_{\mathbf{r},y})^2 + (u_{\mathbf{r}+\hat{y},x}-u_{\mathbf{r},x})^2\right]
\end{split}
\end{equation}
Here, $K_1$ and $K_2$ are the longitudinal and transverse (shear)
stiffnesses. Notably, $K_2$ is proportional to the tension exerted on the Te net by its crystalline environment\cite{EM}. Because Eq.~\eqref{Eq:HamPh} depends only on relative displacements, it is invariant under a uniform shift $u_{\mathbf r,a}\rightarrow u_{\mathbf r,a}+\delta_a$. In the full crystal, this symmetry is weakly broken by interlayer pinning from the surrounding $R$Te blocks, which we neglect here. Here and throughout, we fix $g_\parallel=0.75\,t/a$ and $K_1=0.5\,t/a^2$, and tune $g_{\perp}/g_{\parallel}$ and $K_2/K_1$. Varying $g_{\parallel}$ and $K_1$ shifts the phase boundaries, but the nature of the observed phases is relatively robust over a range of values.

\begin{figure}[t]
\centering
\includegraphics[width=1.0\linewidth]{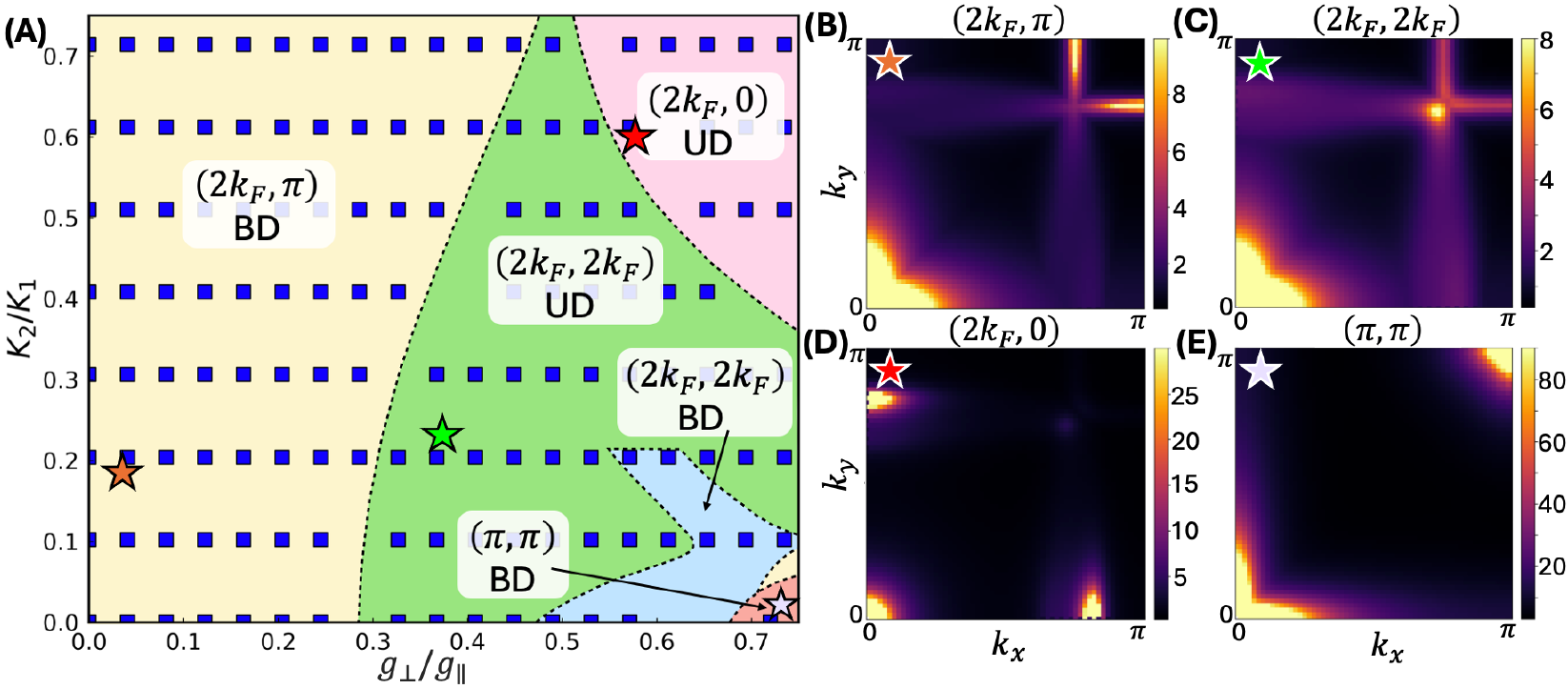}
\caption{\label{fig:LGphase}\textbf{(A)}
Phase diagram showing the nature of the CDW order at $T$ slightly below $T_{c1}$ for fixed $g_\parallel = 0.75\ t/a$ and $K_1 = 0.5\ t/a^2$  as a function of $g_{\perp}/g_{\parallel}$ and $K_2/K_1$. Each region is labeled by the ordering wavevector and by its unidirectional (UD) or
bidirectional (BD) character. \textbf{(B-E)} Representative susceptibility maps at $T=T_{c1}$ for the four starred points in (A) at which ordering types $(2k_F,\pi), (2k_F,2k_F), (2k_F, 0), \&  \ (\pi,\pi)$ occur. Only one quadrant of the Brillouin zone is shown.
}
\end{figure}
\textit{High-temperature CDW instabilities}---We analyze the onset of CDW order by expanding the mean-field (Landau) free energy derived from Eq.~\ref{eq:FullModel} in powers of phonon distortions $u_x$ and $u_y$. The transition temperature $T_{c1}$ is reached when the normal-state phonon susceptibility $\chi_n$ diverges at a finite wavevector $\bm{Q}$. Fourfold rotation and mirror symmetries generally produce several symmetry-related divergent momenta. The quartic terms in the free-energy expansion determine if this leads to unidirectional or multidirectional CDWs. In Fig.~\ref{fig:LGphase}(A) we present the resulting phase diagram, which shows the leading high-temperature CDW instability\cite{SM}.

At large values of $K_2/K_1$, increasing $g_\perp/g_\parallel$ (making the electron--phonon coupling more isotropic) drives the leading instability from a bidirectional CDW with $\mathbf Q=(2k_F,\pi)$ and $(\pi,2k_F)$(the yellow region in Fig.\ref{fig:LGphase}), to a unidirectional
$(2k_F,2k_F)$ CDW (green region), and finally to unidirectional $(2k_F,0)$
order (pink region). The green region closely resembles the experimentally observed high-temperature CDW. At small $K_2/K_1$, the $\bm{Q} = (2k_F, 0)$ regime is absent; instead, large $g_{\perp}/g_{\parallel}$ favors a bidirectional CDW with $\bm{Q} = (2k_F, 2k_F)$ and $(2k_F, -2k_F)$, and a checkerboard $(\pi, \pi)$ CDW. Except for the checkerboard, all ordering wavevectors correspond to one of the Fermi-surface nesting vectors discussed above.

\begin{figure}[t!]
\centering
\includegraphics[width=0.92\linewidth]{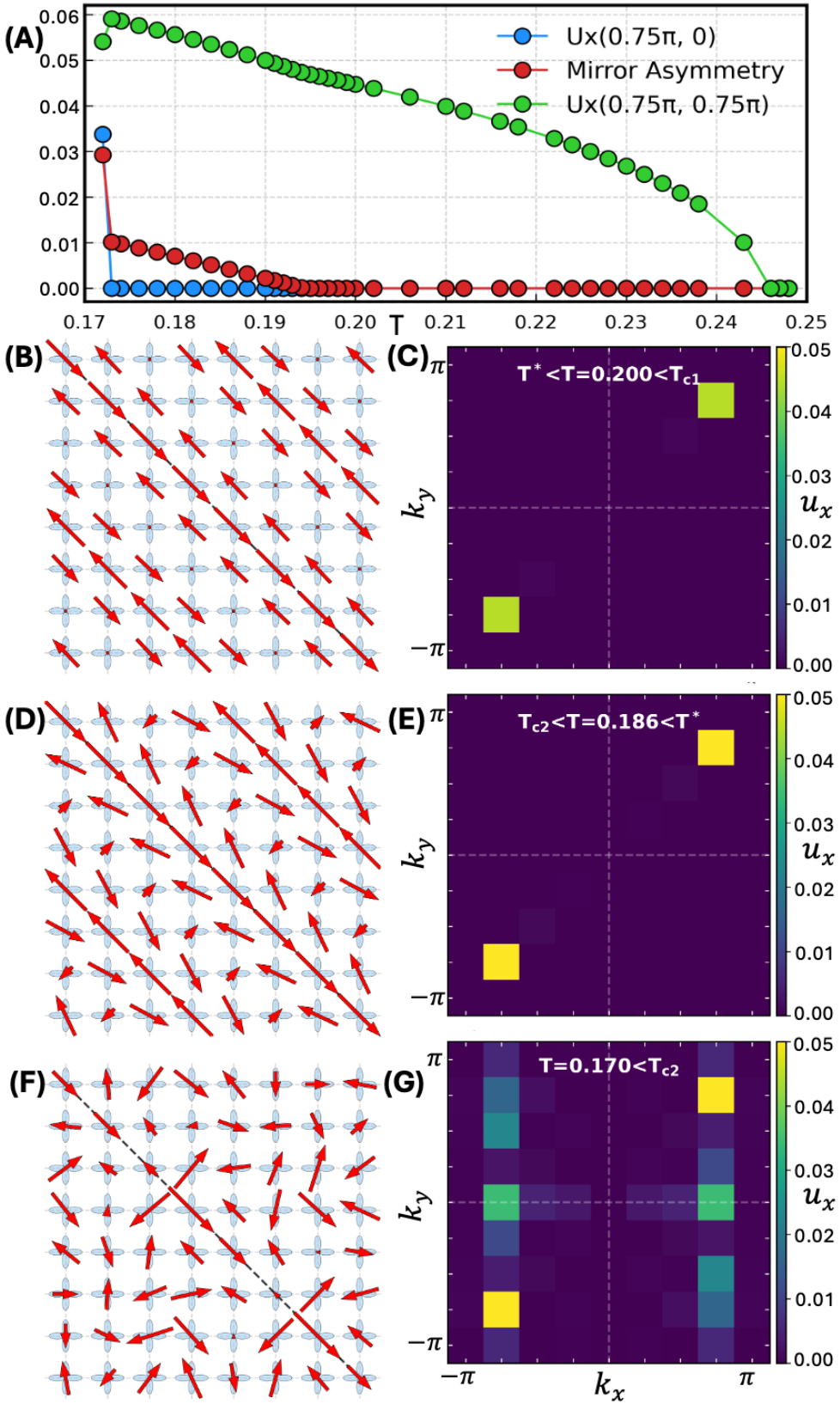}
\caption{\label{fig:CDW_3Transition}\textbf{(A)}Temperature evolution of $|u_x(\mathbf Q)|$ at $\mathbf Q=2\pi(\frac{3}{8},\frac{3}{8})$ and $ 2\pi(\frac{3}{8},0)$, that onset at $T_{c1}$ and $T_{c2}$ respectively, and a glide-asymmetry measure defined as the site-averaged squared mismatch between the optimized profile and its glide-transformed image, marking $T^*$, for $g_\perp/g_\parallel=0.55$ and $K_2/K_1=0.27$ in the green region of Fig.~\ref{fig:LGphase}. \textbf{(B,D,F)} Optimized real-space distortions for $T^*<T<T_{c1}, T_{c2}<T<T^*$, and $T<T_{c2}$, respectively. \textbf{(C,E,G)} Corresponding Fourier maps of $u_x$. For better visibility, distortion vectors in (B,D,F) have been enlarged by factors of 10, 10, and 6.
}
\end{figure}

\textit{Low-temperature evolution of the unidirectional $(2k_F, 2k_F)$ phase}---Having identified an extended parameter regime with unidirectional $(2k_F, 2k_F)$ CDW order, we now analyze its evolution as temperature is lowered. Below $T_{c1}$, the lattice distortions  need not remain small, so a truncated Landau expansion is no longer controlled. We therefore minimize the free energy directly over spatial configurations of $u_{x}$ and $u_{y}$\footnote{As the free energy is invariant under uniform translations, we fix both the sum over all $u_{x}$'s and the sum over all $u_{y}$'s to be zero.}. For numerical tractability, we approximate the incommensurate ordering vector by $\mathbf Q=2\pi(\frac{3}{8},\frac{3}{8})$ which has the highest transition temperature among commensurate approximates $2\pi (\frac{n}{m}, \frac{n}{m})$ with $m\leq 10$ (despite the fact that $\frac{2}{7}$ is closer to the true ordering vector), providing a well-defined 8$\times$8 supercell for optimization.

\begin{figure}[t!]
\centering
\includegraphics[width=1.0\linewidth]{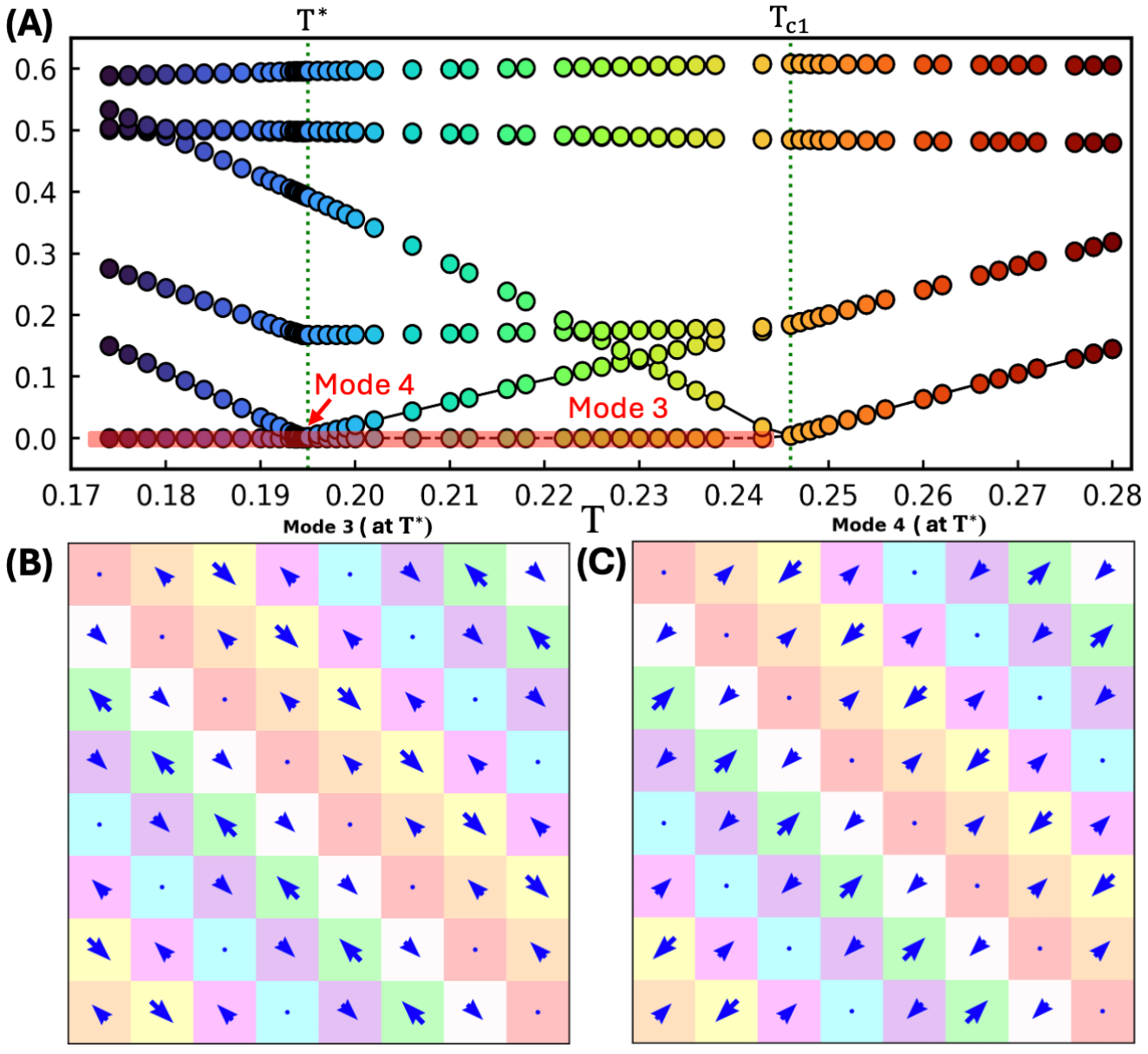}
\caption{\label{fig:phonons}\textbf{(A)} Temperature dependence of the low-energy non-Goldstone spectrum. Notably, the mirror-odd sliding mode (Mode 3, highlighted by the red shading) is anomalously soft from immediately below $T_{c1}$, whereas a secondary soft mode (Mode 4, indicated by the arrow) emerges specifically at $T^*$ and drives the glide symmetry-breaking instability\cite{SM_mirror}. \textbf{(B,C)} Real-space profiles of Mode 3 and 4. Comparison with Fig.~\ref{fig:CDW_3Transition}(B) shows that Mode 3 is odd under $M_{x+y}$. Background colors
label the eight inequivalent site classes of the period-8 supercell; identically colored sites are related by preserved translations and share
the same displacement.
}
\end{figure}

For parameters in the unidirectional $(2k_F, 2k_F)$ region of Fig.~\ref{fig:LGphase}(A), $u_{x}$ and $u_{y}$ develop unidirectional modulations at wavevector $\bm{Q} = 2\pi(\frac{3}{8}, \frac{3}{8})$ slightly below the value of $T_{c1}$ obtained from the Landau expansion, as expected; hereafter we denote this commensurate transition by $T_{c1}$. Just below $T_{c1}$, the CDW has a transverse character [Fig.~\ref{fig:CDW_3Transition}(B)], consistent with experiment\cite{Kogar2020,Maschek2018,Zong2021-2}. The state is invariant under translations perpendicular to $\mathbf Q$ and preserves $M_{x+y}$ and a glide symmetry $M'_{x-y}$, defined as $M_{x-y}$ mirror followed by a half-period translation. Although $M_{x-y}$ itself is broken, $M'_{x-y}$ produces many of the same experimental signatures. The state also breaks the $C_4$, $M_x$ and $M_y$ symmetries of the tetragonal single Te-net model, which are absent in real $R$Te$_3$ crystals.

Although the primary unidirectional CDW forms at $T_{c1}$, the system can still undergo additional low-temperature phase transitions in which one or more residual symmetries are broken. Depending on the value of $K_2/K_1$, we find two distinct symmetry-breaking sequences. We first consider the richer transition sequence shown in Fig.~\ref{fig:CDW_3Transition} for representative parameters $g_\perp/g_\parallel=0.55$ and $K_2/K_1=0.27$. Below the primary CDW onset at $T_{c1}=0.246$, a continuous transition at $T^*=0.194$ breaks the glide symmetry $M'_{x-y}$ without introducing an additional ordering wavevector. A subsequent first-order transition at $T_{c2}=0.172$ produces a bidirectional CDW that breaks the remaining perpendicular translation symmetry and $M'_{x-y}$ while preserving $M_{x+y}$. The bidirectional CDW contains several nontrivial harmonics, including a strong $u_x$ peak at $\pm (2\pi \frac{3}{8},0)$.

\begin{figure}[t]
\centering
\includegraphics[width=1.0\linewidth]{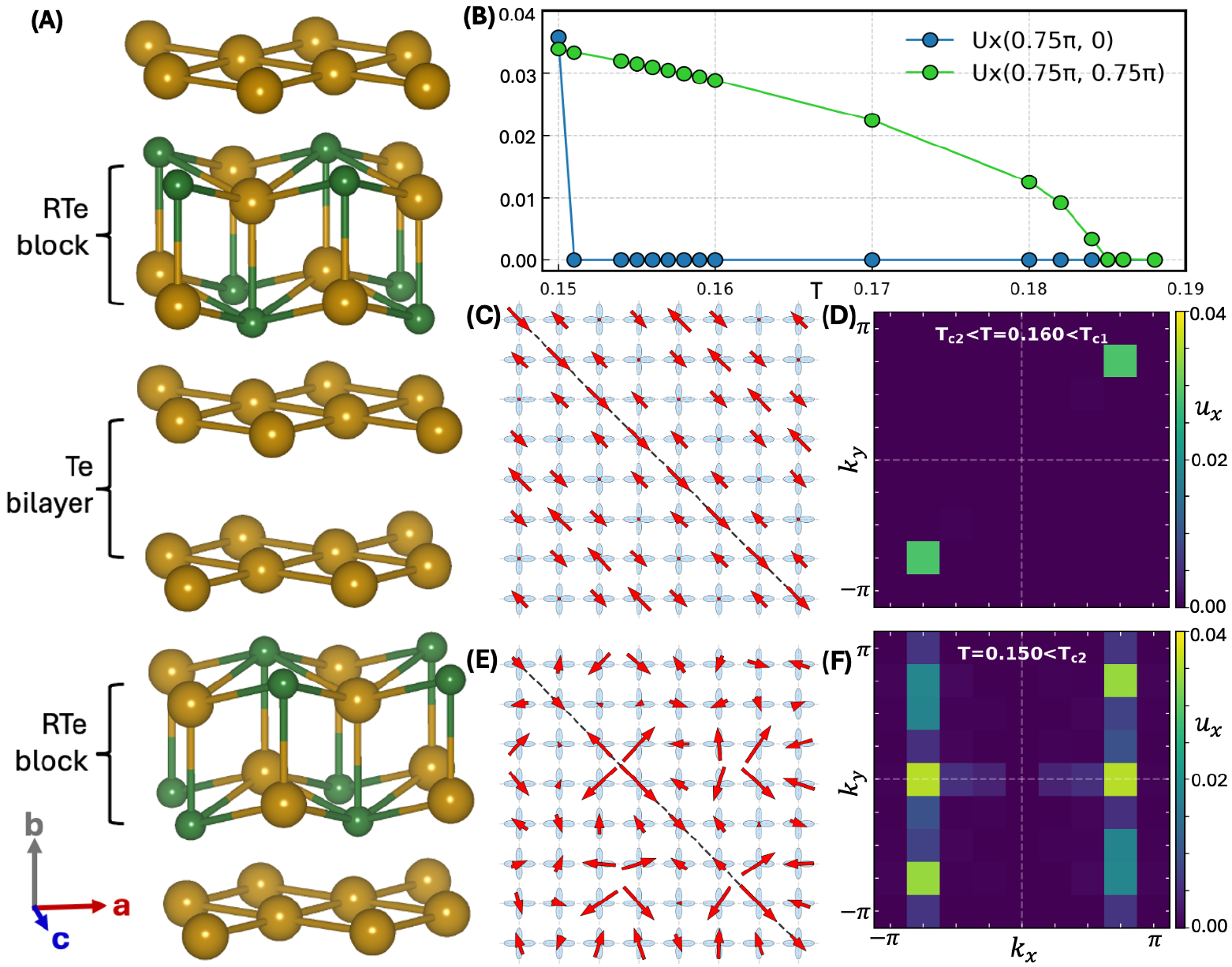}
\caption{\label{fig:CDW_2Transition}\textbf{(A)} Crystal structure of $R$Te$_3$. \textbf{(B)} Temperature dependence of $|u_x(\mathbf{Q})|$ at $2\pi(\frac{3}{8},\frac{3}{8})$ and $ 2\pi(\frac{3}{8},0)$, marking $T_{c1}$ and $T_{c2}$, for $K_2/K_1= 0.35$ and $g_{\perp}/g_{\parallel} = 0.55$ ($g_\parallel=0.75\,t/a, K_1=0.5\,t/a^2$).
\textbf{(C,D)} Optimized real-space distortion and corresponding $u_x$ Fourier
map at $T=0.160$ where the unidirectional $(2k_F, 2k_F)$ CDW is established. \textbf{(E,F)} Analogous plots directly below $T_{c2}$, the second, lower-temperature CDW transition. The Fourier transform of $u_y$ follows from $M_{x+y}$ reflection. For clarity, the distortion vectors in (C,E) are enlarged by factors of 10 and 6, respectively.
}
\end{figure}

We next increase $K_2/K_1$ to $0.35$ while keeping $g_{\perp}/g_{\parallel}=0.55$ fixed. The increased shear stiffness lowers $T_{c1}$ to $0.185$. Additionally, the lower-temperature glide-symmetry breaking and second CDW transitions merge into a single first-order
transition at $T_{c2}=0.151$ (see Fig.~\ref{fig:CDW_2Transition} for detailed differences compared to Fig.~\ref{fig:CDW_3Transition}).

\textit{Phonon spectrum in the CDW phase}---To understand the dynamics of the $(2k_F, 2k_F)$ CDW, we compute the phonon spectrum by evaluating quadratic atomic fluctuations around the optimized $u_{x}$ and $u_{y}$ configuration in the 8$\times$8 commensurate supercell. We restrict our focus to $T>T_{c2}$, as the bidirectional phase found below $T_{c2}$ does not closely resemble experiment, as discussed below.

Accounting for Brillouin-zone backfolding, there are $16 = 2\times 8$ $\Gamma$-point phonon modes above $T_{c2}$, where the factor of $2$ accounts for $x$ and $y$ distortions. Two are gapless uniform-translation modes (Modes 1 and 2) which we omit below. Fig.~\ref{fig:phonons}(A) shows the low-energy portion of the non-Goldstone spectrum using the same parameters as in Fig.~\ref{fig:CDW_3Transition}. For larger values of the shear stiffness (where there are only two transitions), the phonon spectrum is qualitatively similar except for the absence of the Mode 4 softening associated with $T^*$\cite{SM_mirror}. Below, we focus on the two lowest non-Goldstone modes, Modes 3 and 4.

Above $T_{c1} = 0.246$, Modes $3$ and $4$ are degenerate and separated by a sizable gap from the two gapless modes. At $T_{c1}$ these modes soften, driving the CDW transition. However, surprisingly, Mode $3$ remains anomalously soft throughout the CDW phase, at an energy only $\sim 10^{-4}$ above Modes $1$ and $2$---far below the $\mathcal O(\omega_D/10)$ scale expected from simple mode counting\footnote{With an order-ten number of modes spanning the Debye scale, simple mode counting suggests that the lowest non-Goldstone mode should generically lie at an energy of order $\omega_D/10$.}. Mode $4$ reopens a sizable gap below $T_{c1}$, before softens again at $T^*$ and driving the glide symmetry-breaking transition. Below $T^*$, Mode $4$ becomes gapped again.

Inspection of the real-space profile of Mode $3$ below $T_{c1}$ [Fig.~\ref{fig:phonons}(B)] reveals that it is the sliding mode of the unidirectional CDW\cite{SM}. This mode should be gapped in our commensurate approximation, but the gap here is anomalously small, such that the mode resembles a truly gapless Goldstone mode. Moreover, this mode is odd under the $M_{x+y}$ mirror symmetry, whereas the equilibrium CDW is even; thus, pinning this mode breaks the mirror symmetry of the unidirectional CDW.

\textit{Comparison to Experiment}---Our high-temperature $(2k_F, 2k_F)$ instability and its transverse character agree well with observations in $R$Te$_3$ \cite{Maschek2018}. Additionally, we find a nearly gapless phonon mode corresponding to the sliding mode expected for the incommensurate $R$Te$_3$ CDW. While such a mode is strictly gapless in the absence of disorder, its extremely small gap even in our commensurate approximation is surprising. Because the $R$Te$_3$ ordering vector is nearly commensurate, the CDW may consist of locally commensurate regions separated by discommensurations. Contrariwise, the small gap implies that commensurability effects are likely to be weak.

The appearance of a second low-temperature CDW transition is similar to what occurs in $R$Te$_3$ compounds with heavier rare earths ($R$ = Tm, Er, Ho, and Dy). The resulting low-temperature CDW contains two orthogonal ordering vectors of unequal amplitudes and breaks in-plane translational symmetry in both directions, in agreement with experiment. The correspondence
of the detailed harmonic content is less certain, and, contrary to experiment, we find that the $M_{x+y}$ mirror symmetry is unbroken down to the lowest temperature in our model.

While we do not find any phase with both $M_{x\pm y}$ broken, as reported experimentally\cite{Singh2025,Wang2022,freitas2026revealingnaturechargedensity}, the $T^*$ phase found in our model provides a possible route toward such a state: it already has broken glide symmetry $M'_{x-y}$ and an abnormally soft sliding mode. If this mode were further pinned, the resulting state would break all mirror symmetries, although, to our knowledge, no continuous transition analogous to $T^*$ has been reported in $R$Te$_3$.

The absence of such a transition suggests that these compounds more closely resemble the no-$T^*$ regime exemplified by $K_2/K_1=0.35$. However, since the $T^*$ transition is continuous, the Mode $4$ gap should open continuously as $K_2/K_1$ increases, so an analogous soft Mode $4$ may persist for $K_2/K_1$ values somewhat smaller than 0.35. Such a soft mode may be particularly relevant to heavier-$R$ compounds, where the Te–Te bond lengths approach those of elemental Te and the tensile stress on the Te nets is reduced, thus mapping to smaller $K_2/K_1$\cite{EM}. Weak pinning could then induce static components of both Modes 3 and 4, producing the fully symmetry-broken phase. Consequently, traking the evolution of the mirror-symmetry-breaking response across the $R$Te$_3$  series, particular toward lighter $R$, could help clarify its microscopic origin. More broadly, our findings reveal a strong tendency toward complete mirror-symmetry breaking, providing a new theoretical framework for recent experiments in this class of 
materials.

\textit{Acknowledgments}---We gratefully acknowledge important early contributions to this work by Tixuan Tan and Zhaoyu Han. We would also like to thank Chaitanya Murthy, E. Gull, Rafael Fernandes and Kenneth Burch for very helpful discussions and comments. S.Z., I.R.F. and S.A.K. are supported by the US Department of Energy (DOE), Office of Basic Energy Sciences, Materials Sciences and Engineering Division, under Contract No. DE-AC02-76SF00515. JMM is supported by a Leinweber Institute for Theoretical Physics fellowship. Computational work was performed on the Sherlock cluster at Stanford University.

\bibliography{bib}

\widetext
\begin{center}
    {\large\bfseries End Matter}
\end{center}
\narrowtext
\twocolumngrid

\textit{Details of microscopic origin for the phonon sector of the model}---In this section, we derive the phonon sector of the microscopic Hamiltonian introduced in the main text. In general, an electron-phonon system in the adiabatic and tight-binding limit can be described by a Hamiltonian of the form:
\begin{equation}
\begin{split}
\hat{H}&=\sum_{ij} t(|\hat{\br}_{i}-\hat{\br}_j|) \hat{c}^\dagger_i\hat{c}_j \\
&+\sum_{ij} V_\text{nuclei-nuclei}(|\hat{\br}_{i}-\hat{\br}_j|)
+\hat{V}_\text{e-e}+\sum_i \frac{\hat{\bP}^2_i}{2M}
\end{split}
\end{equation}
where $\mathbf{\hat{r}}_i$ is the position operator of the $i$th nucleus, and $\hat{c}_i$ ($\hat{c}_i^\dagger$) is the electron annihilation (creation) operator for an orbital on this atom. We will focus primarily on the effects of the nuclear interaction potential, denoting $V_{\text{nuclei-nuclei}}(r) \equiv V(r)$ for brevity. Assuming that we have solved the semiclassical part of the nuclear energy, $E_{\text{nuclei}}(\{\mathbf{r}_i\})$, by quenching the kinetic energy term, we can identify the global-minimum configuration, denoted by the equilibrium positions $\{\mathbf{R}_i\}$. We can then perform an expansion around this saddle point to obtain an effective model for the lattice fluctuations, namely the phonons. Defining the displacement operators as $\mathbf{\hat{u}}_i = \mathbf{\hat{r}}_i - \mathbf{R}_i$, and using the shorthand notations $\mathbf{R}_{ij} \equiv \mathbf{R}_i - \mathbf{R}_j$, $\mathbf{\hat{u}}_{ij} \equiv \mathbf{\hat{u}}_i - \mathbf{\hat{u}}_j$, and $R = |\mathbf{R}_{ij}|$, the Hamiltonian expanded around this equilibrium configuration is given by:
\begin{equation}
\begin{split}
\hat{H} &= \sum_{ij} \left[ t(R) + t'(|\mathbf{R}_{ij}|) \frac{\mathbf{\hat{u}}_{ij} \cdot \mathbf{R}_{ij}}{|\mathbf{R}_{ij}|} \right] \hat{c}_i^\dagger \hat{c}_j \\[1ex]
&+ \sum_{ij} V'(|\mathbf{R}_{ij}|) \left[ \frac{\mathbf{\hat{u}}_{ij} \cdot \mathbf{R}_{ij}}{|\mathbf{R}_{ij}|} + \frac{(\mathbf{\hat{u}}_{ij} \times \mathbf{R}_{ij})^2}{2|\mathbf{R}_{ij}|^3} \right] \\[1ex]
&+ \sum_{ij} V''(|\mathbf{R}_{ij}|) \frac{(\mathbf{\hat{u}}_{ij} \cdot \mathbf{R}_{ij})^2}{2|\mathbf{R}_{ij}|^2} + \hat{V}_{\text{e-e}} + \sum_i \frac{\mathbf{\hat{P}}_i^2}{2M}
\end{split}
\end{equation}

An important point in this derivation is that $V'(|\mathbf R_{ij}|)$ should not, in general, be set to zero. Although one might expect the equilibrium bond length to coincide with the minimum of an isolated pair potential, this is not the appropriate condition for the Te square net in $R$Te$_3$. As illustrated in Fig.~\ref{fig:CDW_2Transition}(A), each Te layer is embedded in a three-dimensional crystal environment and is subject to stresses from the neighboring rare-earth telluride blocks; the total energy also includes longer-range ionic and electronic contributions. Therefore, the equilibrium configuration is determined by minimizing the total energy of the crystal, rather than by minimizing each individual Te--Te pair potential. As a consequence, the terms proportional to $V'(|\mathbf R_{ij}|)$ in the harmonic expansion should be retained. Physically, they reflect the pressure exerted on the $Te$ nets by the rare-earth spacer layers. The term linear in the displacement, $\mathbf{\hat{u}}_{ij} \cdot \mathbf{R}_{ij}/{|\mathbf{R}_{ij}|}$, cancels in the bulk after summing over symmetry-related bonds in the equilibrium configuration, but the associated quadratic term for transverse distortions remains and contributes to the effective shear stiffness $K_2$ of the phonon sector. This provides a microscopic motivation for keeping both longitudinal and shear elastic terms in the Hamiltonian Eq.~\ref{Eq:HamPh} introduced in the main text.\\
\\\indent\textit{Details of the direct transition without an intermediate phase}---As discussed in the main text, increasing the shear stiffness to $K_2/K_1 = 0.35$ at fixed $g_\perp/g_\parallel = 0.55$ suppresses the intermediate phase. Fig.~\ref{fig:CDW_2Transition}(B)-(F) shows details of the order parameters and CDW configurations for this case.

\end{document}


\renewcommand{\thefigure}{S\arabic{figure}}
\renewcommand{\thetable}{S\arabic{table}}
\setcounter{figure}{0}
\author{Sijia Zhao}
\email{sijiazgl@stanford.edu}
\affiliation{Department of Applied Physics, Stanford University, Stanford, CA 94305, USA}
\affiliation{Stanford Institute for Materials and Energy Sciences, SLAC National Accelerator Laboratory, Menlo Park, CA 94025}
\author{Julian May-Mann}
\affiliation{Department of Physics, Stanford University, Stanford, CA 94305, USA}
\affiliation{Stanford Institute for Materials and Energy Sciences, SLAC National Accelerator Laboratory, Menlo Park, CA 94025}

\author{Ian~R.~Fisher}
\affiliation{Stanford Institute for Materials and Energy Sciences, SLAC National Accelerator Laboratory, Menlo Park, CA 94025}
\affiliation{Department of Applied Physics, Stanford University, Stanford, CA 94305, USA}

\author{Steven~A.~Kivelson}
\email{kivelson@stanford.edu}
\affiliation{Department of Physics, Stanford University, Stanford, CA 94305, USA}
\affiliation{Stanford Institute for Materials and Energy Sciences, SLAC National Accelerator Laboratory, Menlo Park, CA 94025}
\date{\today}

\title{\Large Supplemental Material}
\maketitle

\section{I. Ginzburg-Landau analysis}
\label{app:FreeEnergyDeriv}
In this section, we perform a Ginzburg-Landau (GL) free-energy analysis of the microscopic Hamiltonian introduced in the main text. Our goal is to characterize the leading lattice instability and the nature of the CDW ordered state by deriving the free energy as an expansion in the distortion field. After integrating out the fermionic degrees of freedom, we obtain an effective free-energy functional for the phonon field, whose quadratic and quartic terms determine the onset of the instability and the competition among candidate ordered configurations discussed in the main text. In the following subsections, we present the details of the diagrammatic calculation.

\subsection{1. Second-order susceptibility and the onset of CDW}
The partition function of the interacting electron-phonon system is given by
\begin{equation}
\begin{split}
Z &= \int \mathcal{D}[u]\mathcal{D}[\psi^\dagger, \psi] e^{-S_{\text{ph}}[u] - S_{\text{e}}[\psi^\dagger, \psi] - S_{\text{e-ph}}[\psi^\dagger, \psi, u]}\\[1ex]
\end{split}
\end{equation}
where $S_{\text{ph}}$ is the bare phonon action, and the fermionic part including the electron-phonon coupling can be compactly written in Matsubara frequency and momentum space as
$$S_e + S_{e-ph} = \sum_{n} \sum_{\bm{k},\bm{k}'} \psi^\dagger_{\bm{k},n} \left[ \bm{\mathcal{G}}_0^{-1}(\bm{k}, i\omega_n) \delta_{\bm{k},\bm{k}'} + \mathbf{M}_{\bm{k},\bm{k}'} \right] \psi_{\bm{k}',n}$$
Here, $\psi_{\bm{k},n} = (\psi_{\bm{k},p_x,n}, \psi_{\bm{k},p_y,n})^T$ is the fermion spinor, and the bare inverse Green's function is
$$\bm{\mathcal{G}}_0^{-1}(\bm{k}, i\omega_n) = (i\omega_n + \mu)\mathbf{1} - \mathbf{h}_\mathbf{k}$$
The electron-phonon scattering matrix $\mathbf M_{\mathbf k,\mathbf k'}
=
\sum\limits_{a=x,y} u_{\mathbf q,a}\,
\mathbf\Gamma_a(\mathbf k,\mathbf k')$ transfers momentum $\bm{q} = \bm{k} - \bm{k}'$ and is linear in the phonon displacement fields, where $\mathbf\Gamma_a$ is the electron-phonon vertex for displacement
polarization $a$. Integrating out the Grassmann variables $\psi^\dagger, \psi$, we obtain the effective phonon free energy:
$$\mathcal{F}_{\text{eff}}[u] = \mathcal{F}_{ph}[u] - T\sum_n \text{Tr} \ln \left( \bm{\mathcal{G}}^{-1}_0 + \mathbf{M} \right)$$
where the trace $\text{Tr}$ runs over momentum and orbital indices. To determine the leading instability, we expand the fermion-induced term to second order in the phonon field $u_\mathbf{q}$:
$$\text{Tr} \ln (\bm{\mathcal{G}}_0^{-1} + \mathbf{M}) = \text{Tr} \ln \bm{\mathcal{G}}_0^{-1} - \frac{1}{2}\text{Tr}(\bm{\mathcal{G}}_0 \mathbf{M} \bm{\mathcal{G}}_0 \mathbf{M}) + \mathcal{O}(u^4)$$
where the quadratic term yields the phonon susceptibility  matrix $\chi^{(2)}(\mathbf{q})$, which renormalizes the quadratic phonon kernel. The CDW instability occurs when the lowest eigenvalue of this kernel vanishes at a finite ordering wavevector.

\begin{equation} 
\begin{tikzpicture}[
baseline=(current bounding box.center), 
    thick,
    fermion/.style={draw=black, postaction={decorate},
        decoration={markings,mark=at position .55 with {\arrow[scale=1.5]{latex}}}},
    phonon/.style={decorate, draw=black,
        decoration={snake, amplitude=1.5pt, segment length=6pt}},
    vertex/.style={circle, fill=black, inner sep=1.5pt, minimum size=4pt}
]

\coordinate (V_L) at (-1.0, 0);
\coordinate (V_R) at (1.0, 0);

\coordinate (E_L) at (-2.6, 0);
\coordinate (E_R) at (2.6, 0);

\draw[phonon] (E_L) -- (V_L);
\draw[phonon] (V_R) -- (E_R);

\draw[-latex] ($(E_L)!0.3!(V_L) + (0, 0.25)$) -- ($(E_L)!0.7!(V_L) + (0, 0.25)$);
\node[above] at ($(E_L)!0.5!(V_L) + (0, 0.25)$) {$u(\mathbf{q})$};

\draw[-latex] ($(V_R)!0.3!(E_R) + (0, 0.25)$) -- ($(V_R)!0.7!(E_R) + (0, 0.25)$);
\node[above] at ($(V_R)!0.5!(E_R) + (0, 0.25)$) {$u^*(\mathbf{q})$};

\draw[fermion] (V_L) arc[start angle=180, end angle=0, radius=1.0] 
    node[pos=0.5, above=2pt, font=\small] {$\bm{\mathcal{G}}_0(\mathbf{k}+\mathbf{q}, i\omega_n)$};

\draw[fermion] (V_R) arc[start angle=0, end angle=-180, radius=1.0] 
    node[pos=0.5, below=2pt, font=\small] {$\bm{\mathcal{G}}_0(\mathbf{k}, i\omega_n)$};

\node[vertex] at (V_L) {};
\node[vertex] at (V_R) {};

\node[right=2pt, font=\small] at (V_L) {$\mathbf{\Gamma}$};
\node[left=2pt, font=\small] at (V_R) {$\mathbf{\Gamma}$};

\end{tikzpicture}
\qquad~~~\chi^{(2)}(\bm{q}) = \frac{T}{L^2} \sum_{n,\bm{k}} \text{Tr}\left[ \bm{\mathcal{G}}_0(\bm{k}+\bm{q}, i\omega_n) \mathbf{\Gamma}(\bm{k}+\bm{q},\bm{k}) \bm{\mathcal{G}}_0(\bm{k}, i\omega_n) \mathbf{\Gamma}(\bm{k},\bm{k}+\bm{q}) \right]
\end{equation}

\begin{figure}[h!]
\centering
\includegraphics[width=0.6\linewidth]{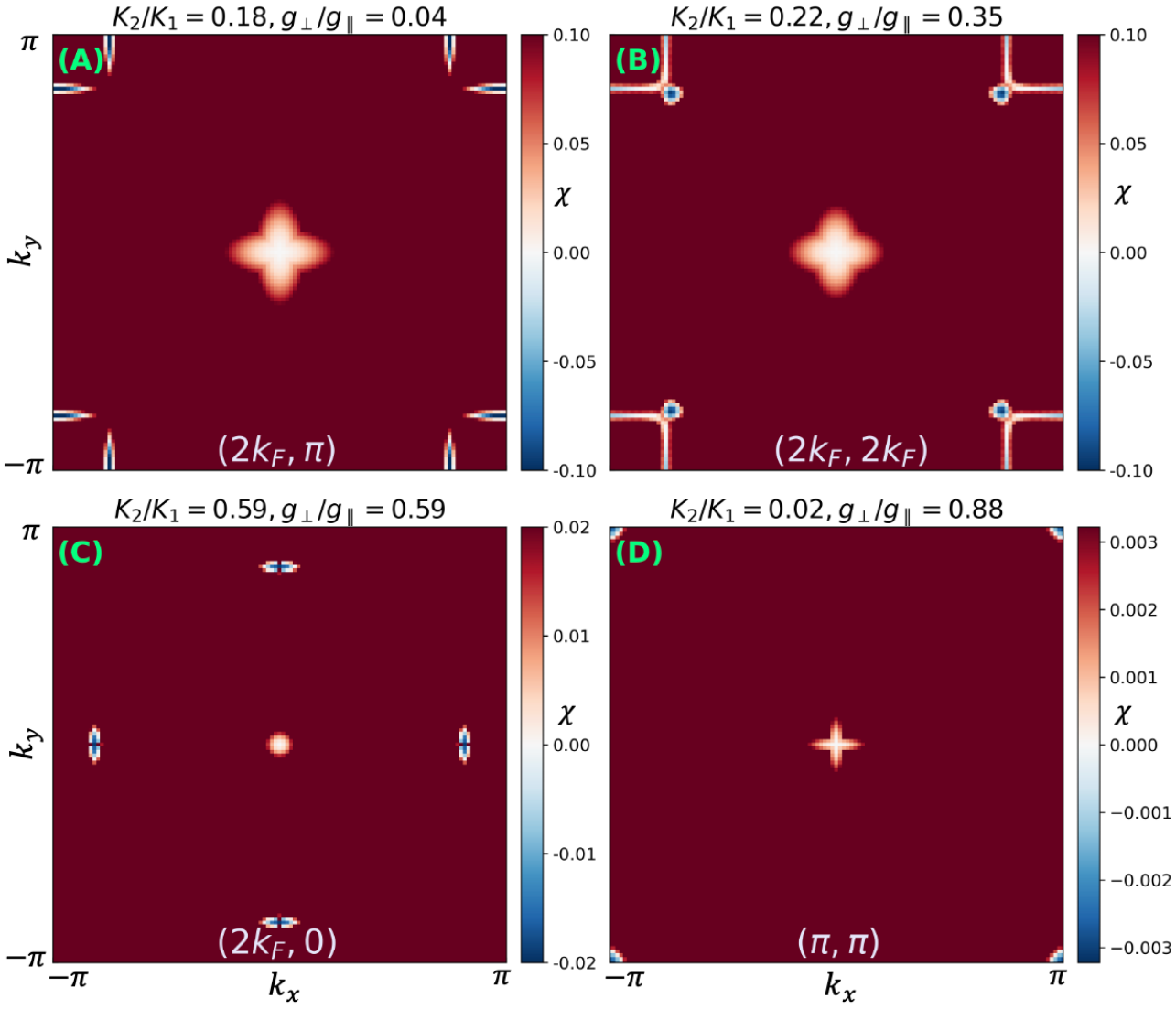}
\caption{\label{SI.1A}Second-order susceptibility for four representative parameter points chosen from the four ordered phases in the main-text phase diagram, evaluated slightly below their respective $T_{c1}$. The ordering vector is indicated at the bottom of each panel. The corresponding parameter values $K_2/K_1$ and $g_{\perp}/g_{\parallel}$ are listed in the panel titles. Here $g_{\parallel}$ and $g_{\perp}$ are the two electron-phonon coupling constants defined in the main tex, and $K_1$ and $K_2$ are the longitudinal and shear elastic constants, respectively. 
}
\end{figure}

\begin{figure}[h!]
\centering
\includegraphics[width=0.86\linewidth]{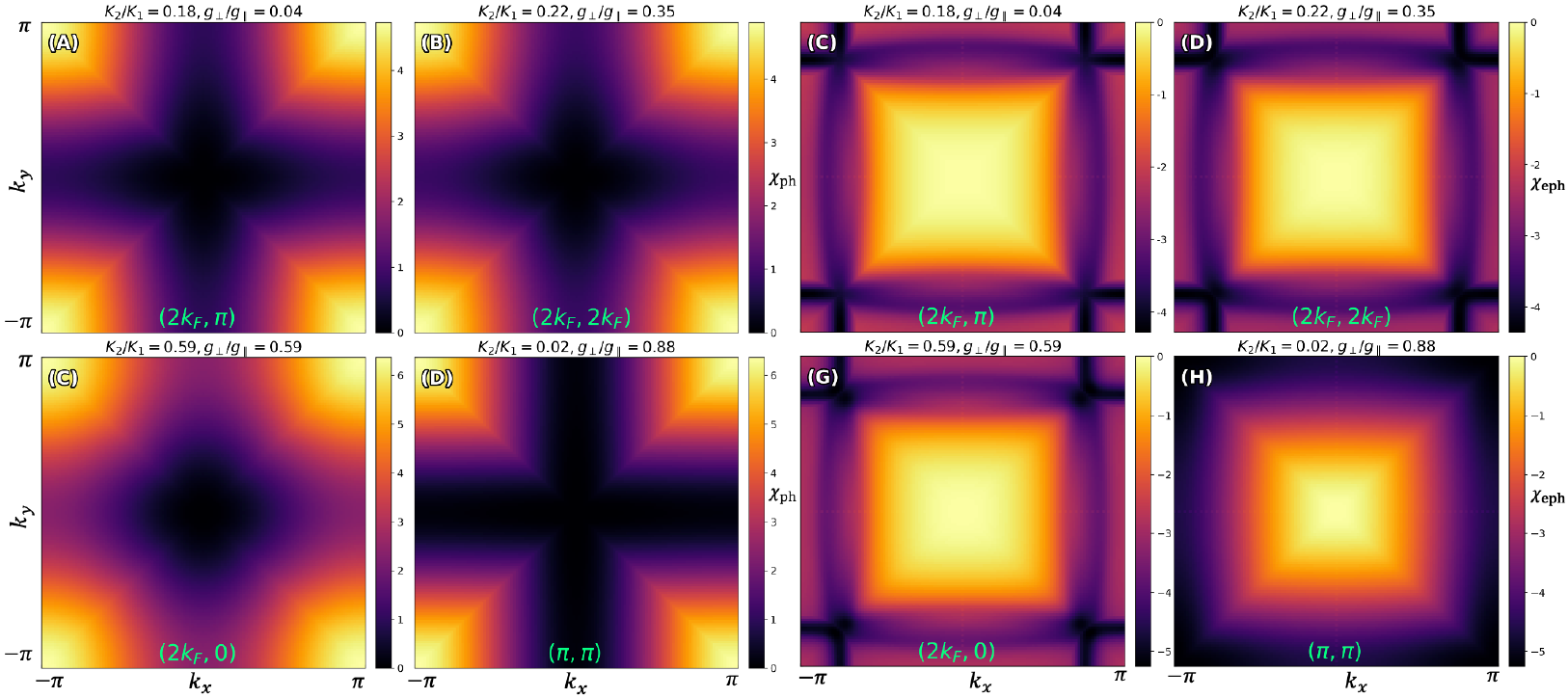}
\caption{\label{SI.1B}Decomposition of the full quadratic kernel shown in Fig.~\ref{SI.1A} into its two constituent parts, evaluated at the same four representative parameter points. The left four panels show the bare lattice term, while the right four panels show the electron-phonon contribution. In each case, the parameter values $K_2/K_1$ and $g_{\perp}/g_{\parallel}$ are the same as those indicated in the corresponding panels of Fig.~\ref{SI.1A}.
}
\end{figure}

At quadratic order, the theory determines which momentum channel becomes unstable first. As shown in Fig.~1 of the main text, the phase diagram contains four distinct ordered phases, characterized by ordering vectors $(2k_F,\pi), (2k_F,2k_F),(2k_F,0)$ and $(\pi,\pi)$, respectively. To illustrate how these instabilities arise from the quadratic theory, we select one representative parameter point from each phase and plot the lowest eigenvalue of the quadratic kernel just below $T_{c1}$, as shown in Fig.~\ref{SI.1A}. Since the temperature is chosen only slightly below the transition, the leading minimum is particularly sharp, allowing the ordering wavevector to be identified unambiguously. The full quadratic kernel consists of the bare lattice term and the electron-phonon contribution; for completeness, we also show these two components separately in Fig.~\ref{SI.1B}.

\begin{figure}[h!]
\centering
\includegraphics[width=0.75\linewidth]{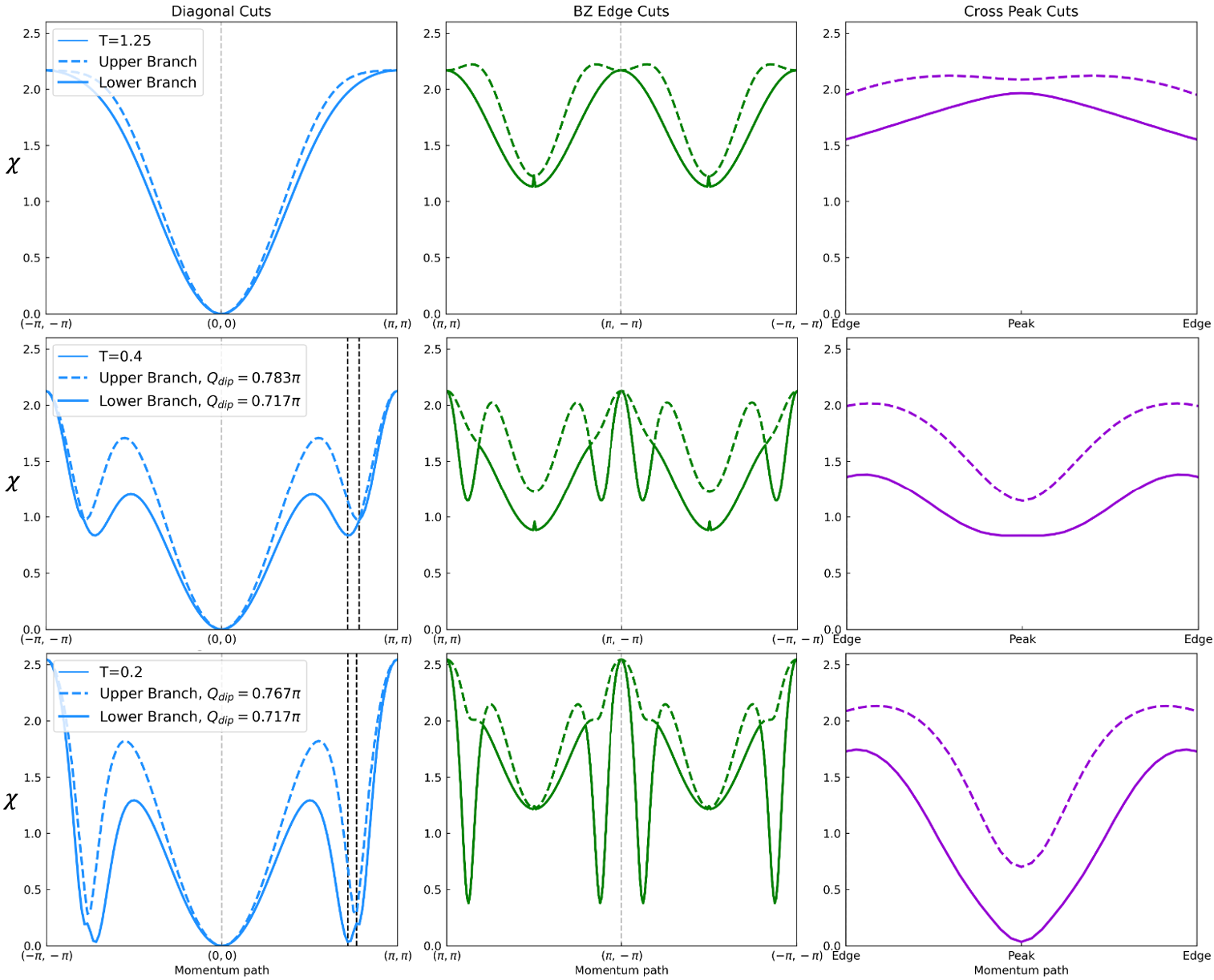}
\caption{\label{SI.2A}High-symmetry momentum cuts of the second-order susceptibility $\chi^{(2)}(\mathbf{q})$ for a representative parameter point in the $(2k_F,2k_F)$ phase, with $K_2/K_1= 0.31$ and $g_{\perp}/g_{\parallel}= 0.49$. The three rows correspond to $T=1.25, 0.40$ and $0.20$, respectively. The first column (blue) shows the diagonal cut along $(-\pi,-\pi)\rightarrow(0,0)\rightarrow(\pi,\pi)$. The second column (green) shows the Brillouin-zone edge cut along $(\pi,\pi)\rightarrow(\pi,-\pi)\rightarrow(-\pi,-\pi)$. The third column (purple) shows the cross-peak cut taken along the diagonal direction perpendicular to the one passing through $\mathbf{Q}$. The dashed and solid curves denote the upper and lower branches, respectively. As indicated in the legends, we also label the momentum positions at which the upper and lower branches first develop soft dips. The transition temperature is $T_{c1}=0.19$, at which the primary peak occurs at $|Q_x| = |Q_y| \approx 0.717\pi$.
}
\end{figure}

\subsection{2. High-symmetry line cuts: two-branch softening of $\chi^{(2)}(\mathbf{q})$}
To further elucidate the structure of the $(2k_F,2k_F)$ instability, we examine several high-symmetry cuts of the second-order susceptibility $\chi^{(2)}(\mathbf{q})$, as shown in Fig.~\ref{SI.2A}. The cuts are taken at a representative parameter point within the $(2k_F,2k_F)$ phase ($K_2/K_1=0.31, g_{\perp}/g_{\parallel} =0.49 $), at temperatures progressively approaching $T_{c1}$ from above. The most notable feature appears in the diagonal cuts: as the temperature is lowered, the lower branch softens at the primary ordering wavevector, but at the same time the upper branch also develops substantial softening at a nearby but distinct momentum.

\subsection{3. Quartic box diagrams and symmetry selection of the CDW order}
To determine how the leading CDW instability selected at quadratic order evolves into a specific ordered state, we expand the fermion-induced contribution to the free energy to fourth order in the phonon fields. The resulting quartic contribution is:
\begin{equation}
\begin{split}
\mathcal{F}^{(4)} &= \frac{T}{4}\sum_{n,\mathbf{k},\{\mathbf{q}_i\}}\text{Tr}_{\text{orb}}\Big[\bm{\mathcal{G}}_0(\mathbf{k})\mathbf{M}(\mathbf{k},\mathbf{k}-\mathbf{q}_1)\bm{\mathcal{G}}_0(\mathbf{k}-\mathbf{q}_1)\mathbf{M}(\mathbf{k}-\mathbf{q}_1,\mathbf{k}-\mathbf{q}_1-\mathbf{q}_2)\bm{\mathcal{G}}_0(\mathbf{k}-\mathbf{q}_1-\mathbf{q}_2)\\[0.5ex]
&~~~~~~~~~~~~~~~~~~~~~~~~~\mathbf{M}(\mathbf{k}-\mathbf{q}_1-\mathbf{q}_2,\mathbf{k}-\mathbf{q}_1-\mathbf{q}_2-\mathbf{q}_3)\bm{\mathcal{G}}_0(\mathbf{k}+\mathbf{q}_4)\mathbf{M}(\mathbf{k}+\mathbf{q}_4,\mathbf{k})\Big]\delta\Big(\sum^4_{i=1}
\mathbf{q}_i-\mathbf{G}\Big)
\end{split}
\end{equation}
Here, $\mathbf{G}$ is a reciprocal lattice vector, and each $\mathbf{M}$ denotes an electron-phonon scattering matrix carrying the appropriate external momentum and form-factor structure as defined previously. 
Restricting the external phonon legs to the primary ordering wavevectors $\mathbf{Q}_1=(2k_F,2k_F), \mathbf{Q}_2=(2k_F,\pi)\text{ and } \mathbf{Q}_3=(2k_F,0)$, together with their corresponding $C_4$-related counterpart $\mathbf{\tilde{Q}}_1,\mathbf{\tilde{Q}}_2 \text{ and } \mathbf{\tilde{Q}}_3$, and combining this quartic contribution with the quadratic term derived above, we obtain the Ginzburg--Landau functional. As an example, for the pair of $C_4$-related wavevectors $\mathbf{Q}_1$ and $\mathbf{\Tilde{Q}}_1$, the free energy takes the form:
\begin{equation}
\begin{split}
\mathcal{F}&=\alpha_1|u_x(\mathbf{Q}_1)|^2+\alpha_2|u_y(\mathbf{Q}_1)|^2+\lambda_1\text{Re}[u^*_x(\mathbf{Q}_1)u_y(\mathbf{Q}_1)]\\[1ex]
&~~+\alpha_2|u_x(\mathbf{\Tilde{Q}}_1)|^2+\alpha_1|u_y(\mathbf{\Tilde{Q}}_1)|^2+\lambda_2\text{Re}[u^*_x(\mathbf{\Tilde{Q}}_1)u_y(\mathbf{\Tilde{Q}}_1)]\\[1ex]
&~~+ \beta_1|u_x(\mathbf{Q}_1)|^4 + \beta_2|u_x(\mathbf{\Tilde{Q}}_1)|^4 + \beta_2|u_y(\mathbf{Q}_1)|^4 + \beta_1|u_y(\mathbf{\Tilde{Q}}_1)|^4 \\[1ex]
&~~+ \gamma_1\Big(|u_x(\mathbf{Q}_1)|^2 |u_x(\mathbf{\Tilde{Q}}_1)|^2 +|u_y(\mathbf{Q}_1)|^2 |u_y(\mathbf{\Tilde{Q}}_1)|^2\Big)\\[1ex]
&~~+\gamma_2\Big(|u_x(\mathbf{Q}_1)|^2|u_y(\mathbf{Q}_1)|^2+|u_x(\mathbf{\Tilde{Q}}_1)|^2|u_y(\mathbf{\Tilde{Q}}_1)|^2\Big)\\[1ex]
&~~+\gamma_3|u_x(\mathbf{Q}_1)|^2|u_y(\mathbf{\Tilde{Q}}_1)|^2+\gamma_4|u_x(\mathbf{\Tilde{Q}}_1)|^2|u_y(\mathbf{Q}_1)|^2\\[1ex]
&~~+\gamma_5\Big(\text{Re}[u_x^2(\mathbf{Q}_1)u_y^{*2}(\mathbf{Q}_1)]+\text{Re}[u_x^2(\mathbf{\Tilde{Q}}_1)u_y^{*2}(\mathbf{\Tilde{Q}}_1)] \Big)\\[1ex]
&~~+\gamma_6\Big(\text{Re}[u_x(\mathbf{Q}_1)u_x(\mathbf{\Tilde{Q}}_1)u_y^{*}(\mathbf{Q}_1)u_y^{*}(\mathbf{\Tilde{Q}}_1)]+ \text{Re}[u_x(\mathbf{Q}_1)u_x^*(\mathbf{\Tilde{Q}}_1)u_y^{*}(\mathbf{Q}_1)u_y(\mathbf{\Tilde{Q}}_1)]\Big)+...
\end{split}
\end{equation}
The quartic coefficients are evaluated diagrammatically from fermionic box diagrams with four external phonon legs.

\begin{tikzpicture}[
    thick,
    fermion/.style={draw=black, postaction={decorate},
        decoration={markings,mark=at position .55 with {\arrow[scale=1.5]{latex}}}},
    phonon/.style={decorate, draw=black,
        decoration={snake, amplitude=1.5pt, segment length=6pt}},
    vertex/.style={circle, fill=black, inner sep=1.5pt, minimum size=4pt}
]

\node at (-3.2, 0) {\LARGE $\gamma_3 :$};

\coordinate (TL) at (-0.8, 0.8);  
\coordinate (TR) at (0.8, 0.8);   
\coordinate (BR) at (0.8, -0.8);  
\coordinate (BL) at (-0.8, -0.8); 

\coordinate (E_TL) at (-1.8, 1.8);
\coordinate (E_TR) at (1.8, 1.8);
\coordinate (E_BR) at (1.8, -1.8);
\coordinate (E_BL) at (-1.8, -1.8);

\draw[phonon] (E_TL) -- (TL);
\draw[phonon] (E_TR) -- (TR);
\draw[phonon] (E_BR) -- (BR);
\draw[phonon] (E_BL) -- (BL);

\draw[-latex] ($(E_TL)!0.3!(TL) + (0.2, 0.2)$) -- ($(E_TL)!0.7!(TL) + (0.2, 0.2)$);
\node[above left] at ($(E_TL) + (0.1,-0.1)$) {$u_x(\mathbf{Q}_1)$};

\draw[-latex] ($(E_TR)!0.3!(TR) + (-0.2, 0.2)$) -- ($(E_TR)!0.7!(TR) + (-0.2, 0.2)$);
\node[above right] at ($(E_TR) + (-0.1,-0.1)$) {$u_y(\tilde{\mathbf{Q}}_1)$};

\draw[-latex] ($(E_BR)!0.3!(BR) + (-0.2, -0.2)$) -- ($(E_BR)!0.7!(BR) + (-0.2, -0.2)$);
\node[below right] at ($(E_BR) + (-0.1,0.1)$) {$u^*_x(\mathbf{Q}_1)$};

\draw[-latex] ($(E_BL)!0.3!(BL) + (0.2, -0.2)$) -- ($(E_BL)!0.7!(BL) + (0.2, -0.2)$);
\node[below left] at ($(E_BL) + (0.1,0.1)$) {$u^*_y(\tilde{\mathbf{Q}}_1)$};

\draw[fermion] (BL) -- (TL) node[pos=0.5, left=2pt, font=\small] {$\bm{\mathcal{G}}_0(\mathbf{k})$};
\draw[fermion] (TL) -- (TR) node[pos=0.5, above=2pt, font=\small] {$\bm{\mathcal{G}}_0(\mathbf{k}+\mathbf{Q}_1)$};
\draw[fermion] (TR) -- (BR) node[pos=0.5, right=2pt, font=\small] {$\bm{\mathcal{G}}_0(\mathbf{k}+\mathbf{Q}_1+\tilde{\mathbf{Q}}_1)$};
\draw[fermion] (BR) -- (BL) node[pos=0.5, below=2pt, font=\small] {$\bm{\mathcal{G}}_0(\mathbf{k}+\tilde{\mathbf{Q}}_1)$};

\node[vertex] at (TL) {};
\node[vertex] at (TR) {};
\node[vertex] at (BR) {};
\node[vertex] at (BL) {};

\begin{scope}[xshift=8.5cm] 

    \node at (-3.2, 0) {\LARGE $\gamma_4 :$};

    \coordinate (TL) at (-0.8, 0.8);  
    \coordinate (TR) at (0.8, 0.8);   
    \coordinate (BR) at (0.8, -0.8);  
    \coordinate (BL) at (-0.8, -0.8); 

    \coordinate (E_TL) at (-1.8, 1.8);
    \coordinate (E_TR) at (1.8, 1.8);
    \coordinate (E_BR) at (1.8, -1.8);
    \coordinate (E_BL) at (-1.8, -1.8);

    \draw[phonon] (E_TL) -- (TL);
    \draw[phonon] (E_TR) -- (TR);
    \draw[phonon] (E_BR) -- (BR);
    \draw[phonon] (E_BL) -- (BL);

    \draw[-latex] ($(E_TL)!0.3!(TL) + (0.2, 0.2)$) -- ($(E_TL)!0.7!(TL) + (0.2, 0.2)$);
    \node[above left] at ($(E_TL) + (0.1,-0.1)$) {$u_y(\mathbf{Q}_1)$};

    \draw[-latex] ($(E_TR)!0.3!(TR) + (-0.2, 0.2)$) -- ($(E_TR)!0.7!(TR) + (-0.2, 0.2)$);
    \node[above right] at ($(E_TR) + (-0.1,-0.1)$) {$u_x(\tilde{\mathbf{Q}}_1)$};

    \draw[-latex] ($(E_BR)!0.3!(BR) + (-0.2, -0.2)$) -- ($(E_BR)!0.7!(BR) + (-0.2, -0.2)$);
    \node[below right] at ($(E_BR) + (-0.1,0.1)$) {$u^*_y(\mathbf{Q}_1)$};

    \draw[-latex] ($(E_BL)!0.3!(BL) + (0.2, -0.2)$) -- ($(E_BL)!0.7!(BL) + (0.2, -0.2)$);
    \node[below left] at ($(E_BL) + (0.1,0.1)$) {$u^*_x(\tilde{\mathbf{Q}}_1)$};

    \draw[fermion] (BL) -- (TL) node[pos=0.5, left=2pt, font=\small] {$\bm{\mathcal{G}}_0(\mathbf{k})$};
    \draw[fermion] (TL) -- (TR) node[pos=0.5, above=2pt, font=\small] {$\bm{\mathcal{G}}_0(\mathbf{k}+\mathbf{Q}_1)$};
    \draw[fermion] (TR) -- (BR) node[pos=0.5, right=2pt, font=\small] {$\bm{\mathcal{G}}_0(\mathbf{k}+\mathbf{Q}_1+\tilde{\mathbf{Q}}_1)$};
    \draw[fermion] (BR) -- (BL) node[pos=0.5, below=2pt, font=\small] {$\bm{\mathcal{G}}_0(\mathbf{k}+\tilde{\mathbf{Q}}_1)$};

    \node[vertex] at (TL) {};
    \node[vertex] at (TR) {};
    \node[vertex] at (BR) {};
    \node[vertex] at (BL) {};

\end{scope}
\end{tikzpicture}

\begin{tikzpicture}[
    thick,
    fermion/.style={draw=black, postaction={decorate},
        decoration={markings,mark=at position .55 with {\arrow[scale=1.5]{latex}}}},
    phonon/.style={decorate, draw=black,
        decoration={snake, amplitude=1.5pt, segment length=6pt}},
    vertex/.style={circle, fill=black, inner sep=1.5pt, minimum size=4pt}
]


\node at (-3.2, 0) {\LARGE $\gamma_5 :$};

\coordinate (TL) at (-0.8, 0.8);  
\coordinate (TR) at (0.8, 0.8);   
\coordinate (BR) at (0.8, -0.8);  
\coordinate (BL) at (-0.8, -0.8); 

\coordinate (E_TL) at (-1.8, 1.8);
\coordinate (E_TR) at (1.8, 1.8);
\coordinate (E_BR) at (1.8, -1.8);
\coordinate (E_BL) at (-1.8, -1.8);

\draw[phonon] (E_TL) -- (TL);
\draw[phonon] (E_TR) -- (TR);
\draw[phonon] (E_BR) -- (BR);
\draw[phonon] (E_BL) -- (BL);

\draw[-latex] ($(E_TL)!0.3!(TL) + (0.2, 0.2)$) -- ($(E_TL)!0.7!(TL) + (0.2, 0.2)$);
\node[above left] at ($(E_TL) + (0.1,-0.1)$) {$u_x(\mathbf{Q}_1)$};

\draw[-latex] ($(E_TR)!0.3!(TR) + (-0.2, 0.2)$) -- ($(E_TR)!0.7!(TR) + (-0.2, 0.2)$);
\node[above right] at ($(E_TR) + (-0.1,-0.1)$) {$u_x(\mathbf{Q}_1)$};

\draw[-latex] ($(E_BR)!0.3!(BR) + (-0.2, -0.2)$) -- ($(E_BR)!0.7!(BR) + (-0.2, -0.2)$);
\node[below right] at ($(E_BR) + (-0.1,0.1)$) {$u^*_y(\mathbf{Q}_1)$};

\draw[-latex] ($(E_BL)!0.3!(BL) + (0.2, -0.2)$) -- ($(E_BL)!0.7!(BL) + (0.2, -0.2)$);
\node[below left] at ($(E_BL) + (0.1,0.1)$) {$u^*_y(\mathbf{Q}_1)$};

\draw[fermion] (BL) -- (TL) node[pos=0.5, left=2pt, font=\small] {$\bm{\mathcal{G}}_0(\mathbf{k})$};
\draw[fermion] (TL) -- (TR) node[pos=0.5, above=2pt, font=\small] {$\bm{\mathcal{G}}_0(\mathbf{k}+\mathbf{Q}_1)$};
\draw[fermion] (TR) -- (BR) node[pos=0.5, right=2pt, font=\small] {$\bm{\mathcal{G}}_0(\mathbf{k}+2\mathbf{Q}_1)$};
\draw[fermion] (BR) -- (BL) node[pos=0.5, below=2pt, font=\small] {$\bm{\mathcal{G}}_0(\mathbf{k}+\mathbf{Q}_1)$};

\node[vertex] at (TL) {};
\node[vertex] at (TR) {};
\node[vertex] at (BR) {};
\node[vertex] at (BL) {};

\begin{scope}[xshift=8.5cm] 

    \node at (-3.2, 0) {\LARGE $\gamma_6 :$};

    \coordinate (TL) at (-0.8, 0.8);  
    \coordinate (TR) at (0.8, 0.8);   
    \coordinate (BR) at (0.8, -0.8);  
    \coordinate (BL) at (-0.8, -0.8); 

    \coordinate (E_TL) at (-1.8, 1.8);
    \coordinate (E_TR) at (1.8, 1.8);
    \coordinate (E_BR) at (1.8, -1.8);
    \coordinate (E_BL) at (-1.8, -1.8);

    \draw[phonon] (E_TL) -- (TL);
    \draw[phonon] (E_TR) -- (TR);
    \draw[phonon] (E_BR) -- (BR);
    \draw[phonon] (E_BL) -- (BL);

    \draw[-latex] ($(E_TL)!0.3!(TL) + (0.2, 0.2)$) -- ($(E_TL)!0.7!(TL) + (0.2, 0.2)$);
    \node[above left] at ($(E_TL) + (0.1,-0.1)$) {$u_x(\mathbf{Q}_1)$};

    \draw[-latex] ($(E_TR)!0.3!(TR) + (-0.2, 0.2)$) -- ($(E_TR)!0.7!(TR) + (-0.2, 0.2)$);
    \node[above right] at ($(E_TR) + (-0.1,-0.1)$) {$u_x(\tilde{\mathbf{Q}}_1)$};

    \draw[-latex] ($(E_BR)!0.3!(BR) + (-0.2, -0.2)$) -- ($(E_BR)!0.7!(BR) + (-0.2, -0.2)$);
    \node[below right] at ($(E_BR) + (-0.1,0.1)$) {$u^*_y(\mathbf{Q}_1)$};

    \draw[-latex] ($(E_BL)!0.3!(BL) + (0.2, -0.2)$) -- ($(E_BL)!0.7!(BL) + (0.2, -0.2)$);
    \node[below left] at ($(E_BL) + (0.1,0.1)$) {$u^*_y(\tilde{\mathbf{Q}}_1)$};

    \draw[fermion] (BL) -- (TL) node[pos=0.5, left=2pt, font=\small] {$\bm{\mathcal{G}}_0(\mathbf{k})$};
    \draw[fermion] (TL) -- (TR) node[pos=0.5, above=2pt, font=\small] {$\bm{\mathcal{G}}_0(\mathbf{k}+\mathbf{Q}_1)$};
    \draw[fermion] (TR) -- (BR) node[pos=0.5, right=2pt, font=\small] {$\bm{\mathcal{G}}_0(\mathbf{k}+\mathbf{Q}_1+\tilde{\mathbf{Q}}_1)$};
    \draw[fermion] (BR) -- (BL) node[pos=0.5, below=2pt, font=\small] {$\bm{\mathcal{G}}_0(\mathbf{k}+\tilde{\mathbf{Q}}_1)$};

    \node[vertex] at (TL) {};
    \node[vertex] at (TR) {};
    \node[vertex] at (BR) {};
    \node[vertex] at (BL) {};

\end{scope}
\end{tikzpicture}

\clearpage
Based on the sign and magnitude of the quartic coefficients $\gamma_i$, we find that over most of the $\mathbf{Q}_1$-ordered phase, $u_x$ and $u_y$ exhibit unidirectional behavior: both components favor the same ordering vector, with either $u_{x,y}(\mathbf{Q}_1)$ or $u_{x,y}(\mathbf{\Tilde{Q}}_1)$ nonzero. We refer to this as the ``$(2k_F,2k_F)  ~\text{UD}$'' order (green region in Fig.~2(A) of the main text). For completeness, we also characterize the other two types of ordered phases. In the $\mathbf{Q}_2$  region, the system becomes bidirectional, with $u_{x}$ and $u_{y}$ favoring different ordering vectors; that is, either $u_{x}(\mathbf{Q}_2)$ and $u_{y}(\mathbf{\Tilde{Q}}_2)$ are nonzero, or vice versa. We refer to this as the ``$(2k_F,\pi)  ~\text{BD}$'' order (yellow region in Fig.~2(A)). And in the $\mathbf{Q}_3$-ordered region, only one component of the order parameter is allowed to be nonzero: either $u_{x}(\mathbf{Q}_3)\neq0$ or $u_{y}(\mathbf{\Tilde{Q}}_3)\neq0$, which is denoted as the ``$(2k_F,0)  ~\text{UD}$'' type of order. Overall, the spontaneous breaking of $C_4$ rotational symmetry between $\mathbf{Q}_1$ and $\mathbf{\Tilde{Q}_1}$ is robustly established within our model, setting the stage for the investigation of macroscopic mirror symmetries breaking.

\section{II. Real-space optimization}
To complement the Ginzburg-Landau analysis above, we also determine the ordered state directly in real space through free-energy minimization. Guided by the ordering wave vector identified from the susceptibility at $T_{c1}$, we construct a commensurate approximation to the dominant ordering tendency. In particular, since the peak occurs near $Q\approx 2\pi\times 3/8$, we adopt an $8\times 8$ real-space unit cell as the minimal commensurate supercell. Within this supercell, we perform an unrestricted minimization over all $2L_xL_y=128$ components of the displacement field, without imposing a single-$Q$ ansatz or any point-group symmetry constraint. For each trial configuration, we construct and diagonalize the full two-orbital Bloch Hamiltonian throughout the reduced Brillouin zone, determine the chemical potential self-consistently to maintain the prescribed filling, and evaluate the canonical free energy including the elastic contribution. After projecting out the uniform-translation mode, the resulting nonlinear free-energy functional is minimized using a quasi-Newton BFGS algorithm.

To assess the sensitivity to initial conditions and guard against metastable solutions, we repeat the optimization for every parameter set using three qualitatively distinct initial configurations, including weak- and strong-amplitude random distortions and a symmetry-informed configuration constructed from the soft mode of the quadratic Ginzburg--Landau theory. The independently initialized runs converge to consistent ordered states.  This procedure retains the full nonlinear dependence of the electronic free energy on the lattice distortion and yields robust real-space configurations that directly reveal the symmetry evolution of the CDW order.

\subsection{1. Thermal evolution and symmetry-distinct phases}
In the main text, we highlighted three representative temperatures, $T_{c1}, T^*$ and $T_{c2}$, to illustrate the evolution of the optimized real-space configuration. Here, we present a more complete picture by showing the full thermal evolution of the configuration as the temperature is lowered from $T_{c1}$ to $T_{c2}$. This allows us to track how the ordered pattern develops below the primary instability, how its symmetry evolve through the intermediate regime, and how the system eventually approaches the low-temperature phase.

\begin{figure}[h!]
\centering
\includegraphics[width=1.0\linewidth]{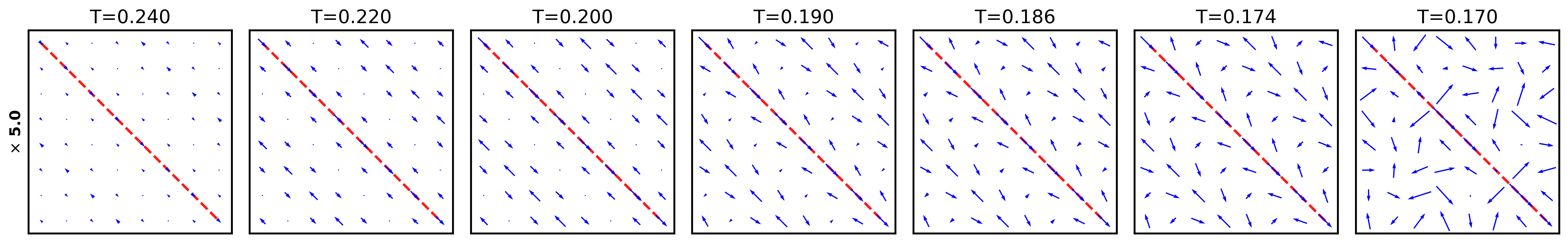}
\caption{\label{SII.1A}
Thermal evolution of the optimized real-space configuration. The panels are arranged from left to right in order of decreasing temperature, spanning the interval from $T_{c1}$ to $T_{c2}$. In the main text, only the three representative configurations are shown; here we present the intermediate configurations as well to illustrate the full evolution of the ordered pattern and its symmetry character through the two transitions. Parameters are the same as in Fig. 3 of the main text, $K_2/K_1= 0.27$ and $g_{\perp}/g_{\parallel}= 0.55$. For visual clarity, the displacement vectors are enlarged by a factor of 5.
}
\end{figure}

\subsection{2. Charge distribution}
As a complementary real-space diagnostic, we also examine the charge distribution at $T_{c1}$ along a horizontal cut through the phase diagram, providing a direct view of how the electronic density rearranges in the optimized state and evolves across nearby parameter regimes. 

\begin{figure}[h!]
\centering
\includegraphics[width=1.0\linewidth]{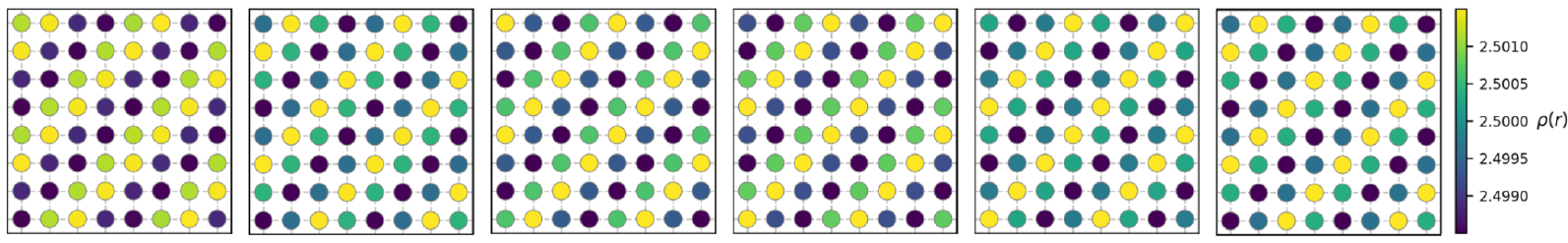}
\caption{\label{SII.2A}Site-resolved charge distribution
$\rho(\mathbf{r})$ just below $T_{c1}$ along the horizontal cut through the phase diagram at fixed $K_2/K_1= 0.31$, for $g_{\perp}/g_{\parallel}=0.49,0.51,0.53,0.55,0.57,$ and $0.59$.
}
\end{figure}

\subsection{3. Extended Data}
\begin{figure}[h!]
\centering
\includegraphics[width=0.88\linewidth]{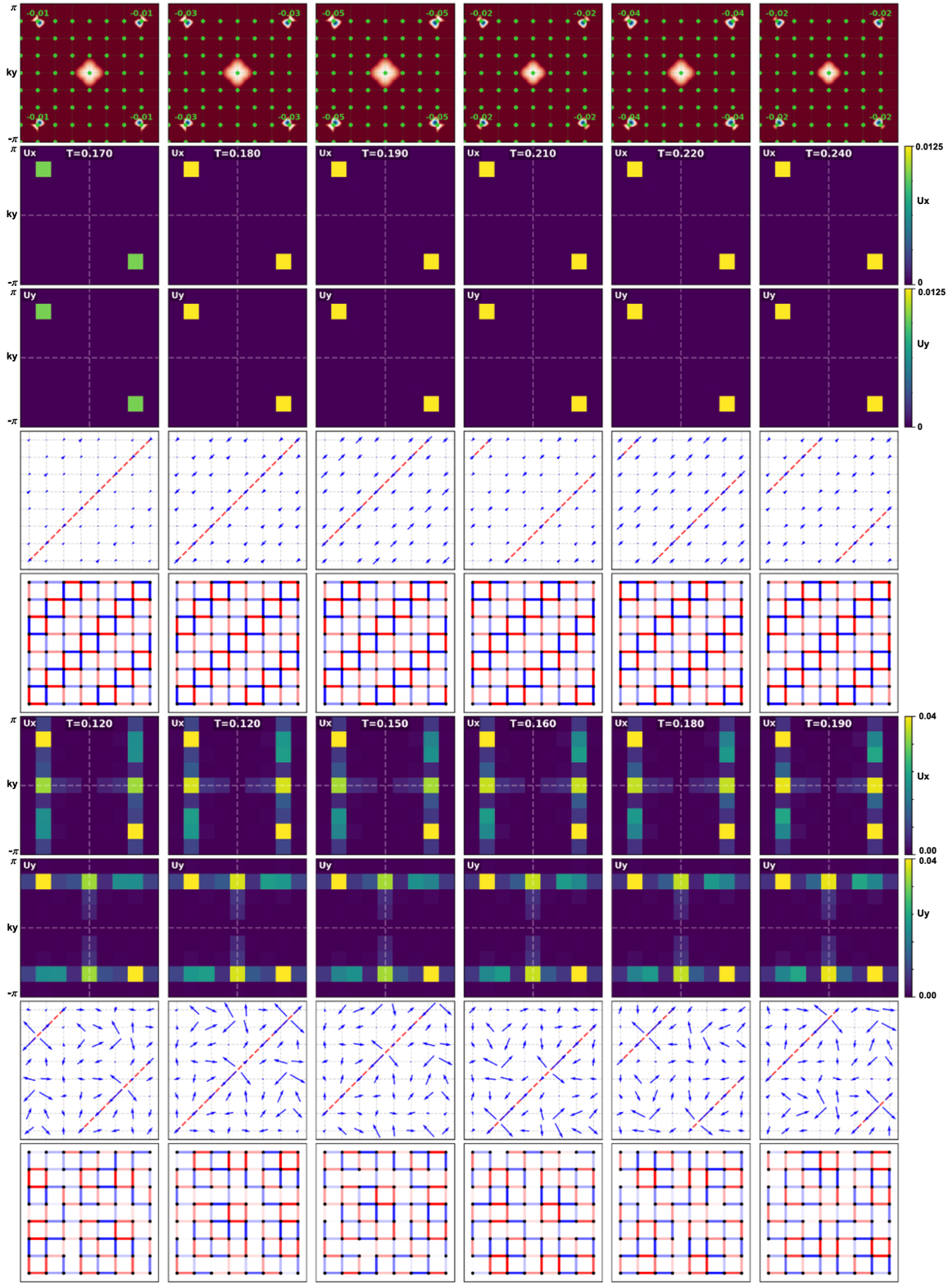}
\caption{\label{SII.3A}Extended data along a horizontal cut of the phase diagram at fixed $K_2/K_1= 0.31$. From left to right, the six columns correspond to $g_{\perp}/g_{\parallel}=0.49,0.51,0.53,0.55,0.57$ and $0.59$, respectively. For each column, the first row shows the second-order susceptibility together with the value on the nearest commensurate point of the $8\times 8$ momentum grid. The second and third rows show the Fourier transforms of $u_x$ and $u_y$ at $T_{c1}$, respectively, highlighting the dominant peak structure. The fourth row displays the optimized real-space distortion pattern at $T_{c1}$, with the red dashed line indicating the corresponding mirror axis. The fifth row shows the same $T_{c1}$ configuration in a bond representation: red bonds indicate locally elongated bonds, blue bonds indicate contracted bonds, and the color intensity encodes the magnitude of the bond-length change. The sixth and seventh rows show the Fourier transforms of $u_x$ and $u_y$ at $T_{c2}$, respectively, while the eighth and ninth rows display the corresponding real-space plots and bond representations. The value of $T_{c1}$ and $T_{c2}$ for each parameter set are indicated in the $u_x$ Fourier-transform panels.
}
\end{figure}

To place the main-text parameter choice in a broader context, we present extended data along the horizontal cut at fixed $K_2/K_1= 0.31$, with $g_{\perp}/g_{\parallel}$ ranging from 0.49 to 0.59. The six columns in Fig.~\ref{SII.3A} correspond, from left to right, to $g_{\perp}/g_{\parallel}=0.49, 0.51, 0.53, 0.55, 0.57$, and $0.59$. This allows a direct comparison between the main-text parameter point and its neighboring regimes.

For each parameter set, we show the susceptibility just below $T_{c1}$ and the value on the nearest commensurate $8\times 8$ momentum grid, together with the optimized real-space and Fourier-space structures at both $T_{c1}$ and $T_{c2}$. At $T_{c1}$, the Fourier components of $u_x$ and $u_y$ corroborate the dominant ordering wave vector selected by the quadratic instability, while the corresponding real-space arrow plot and bond representation visualize the distortion pattern and its associated mirror axis.

\section{III. Mirror-channel response: susceptibility and collective modes}
To characterize the mirror-channel response near the onset state, we analyze the $\Gamma$-point collective modes of the optimized
configuration just below $T_{c1}$ configuration by constructing the free-energy Hessian with respect to the inequivalent-site displacement fields. The corresponding eigenvalues quantify the stiffness of the free energy in different fluctuation channels. In particular, we identify a low-lying mode that is odd under reflection about the mirror axis as shown in Fig.~3 in the main text. Within the quadratic approximation, the inverse stiffness of this mode determines the associated mirror susceptibility. To make this relation explicit, we expand the free energy variation about the optimized configuration $\mathbf{u}^{(0)}$:
\begin{equation}
\begin{split}
F[\mathbf{u}] = F[\mathbf{u}^{(0)}] + \frac{1}{2}\delta \mathbf{u}^T H \delta \mathbf{u} + ...
\end{split}
\end{equation}
where $\delta \mathbf{u}=\mathbf{u} - \mathbf{u}^{(0)}$ and H is the Hessian matrix. Let $v_{M}$ denote the low-lying mirror-odd eigenmode of $\mathbf H$, satisfying $\mathbf H \mathbf v_M=\lambda_M \mathbf v_M$ and
$\mathbf v_M^T \mathbf v_M=1$. Restricting to fluctuations along this channel, $\delta \mathbf{u}=\phi_M \mathbf v_M$, gives
\begin{equation}
\begin{split}
F = F_0 + \frac{\lambda_M}{2}\phi^2_M 
\end{split}
\end{equation}
Introducing a conjugate field $h_M$ coupled linearly to the mirror-odd amplitude $\phi_M$ gives
\begin{equation}
\begin{split}
F = F_0 + \frac{\lambda_M}{2}\phi^2_M - h_M \phi_M
\end{split}
\end{equation}
Minimizing with respect to $\phi_M$ yields $\phi_M = h_M/\lambda_M$. Thus, within the quadratic approximation, the mirror-channel susceptibility is 
\begin{equation}
\begin{split}
\chi_M = \frac{\partial \phi_M}{\partial h_M}\Big|_{h_M\rightarrow 0} = \frac{1}{\lambda_M}
\end{split}
\end{equation}
The anomalous softening of this collective mode therefore signals a strongly enhanced mirror-channel response near $T_{c1}$.

\subsection{1. Collective-mode spectrum} 
\begin{figure}[t!]
\centering
\includegraphics[width=1.0\linewidth]{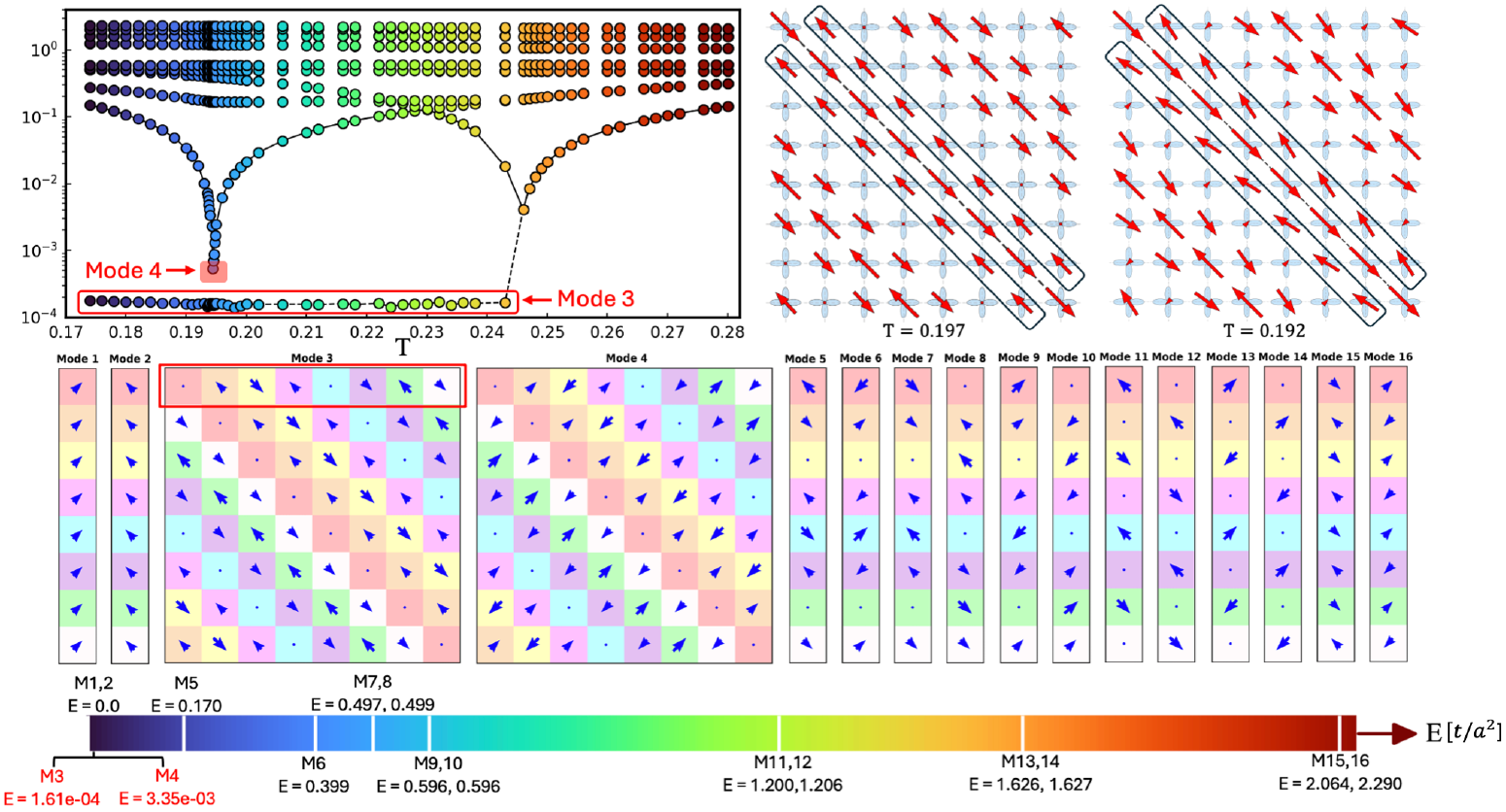}
\caption{\label{fig:SM_phonons}Temperature evolution and real-space signatures of the collective modes for the same parameter point as in Fig.~3 in the main text. This figure provides the extended mode analysis corresponding to Fig.~4 of the main text. \textbf{Top Left:} Temperature dependence of the energies of the 14 non-Goldstone collective modes, shown on a logarithmic vertical scale. Notably, the mirror-odd sliding mode (Mode 3) persists as anomalously soft from immediately below $T_{c1}$, whereas a secondary soft mode (Mode 4) emerges specifically at the $T^*$ phase transition. \textbf{Top Right:} Real-space configurations in the immediate vicinity of $T^*$, at $T=0.197$ (above $T^*$) and $T=0.192$ (below $T^*$). The black boxes highlight a representative diagonal to illustrate the structural transition: the comparison shows that the change across $T^*$ is predominantly along the emergent soft Mode 4. \textbf{Bottom:} Eigenvectors of the 16 $\Gamma$-point collective modes at $T^*$, obtained by allowing two independent displacement components for 8 inequivalent sites. Modes 3, highlighted by its red energy label and plotted on the full $8\times 8$ patch, is the lowest non-Goldstone mode. Comparing it with the CDW configuration (Fig.~3 (B) of the main text), one sees that it is odd under reflection about the mirror axis (the $y=-x$ diagonal). The remaining modes are shown only within the primary supercell, indicated by the red box in the Mode 3 plot. }
\end{figure}

In the main text, we show the low-energy portion of the
14 non-Goldstone $\Gamma$-point modes on a linear scale, using a restricted energy window chosen to resolve the gap closings at $T_{c1}$ and $T^*$. We also display the real-space profiles of Modes 3 and 4. Here, Fig.~\ref{fig:SM_phonons} presents further details of the complete mode structure. The upper-left panel
shows all 14 non-Goldstone modes on a logarithmic scale, while the lower panel displays the corresponding eigenvectors. The upper-right panel compares the optimized configurations immediately above and below $T^*$. Their difference is dominated by Mode 4: the configuration below $T^*$ is well approximated by adding a finite Mode-4 displacement to the configuration above $T^*$, directly identifying Mode 4 with the glide-symmetry-breaking instability.

\subsection{2. The sliding mode}
To elucidate the character of Mode 3, consider a unidirectional CDW
distortion
\begin{equation}
    \mathbf u(\mathbf r)
    =\mathbf u_0\cos(\mathbf Q\cdot\mathbf r+\phi).
\end{equation}
A small translation of the CDW by $\delta\mathbf r$ changes the
distortion by
\begin{equation}
    \delta\mathbf u(\mathbf r)
    \simeq
    \mathbf u_0\sin(\mathbf Q\cdot\mathbf r+\phi)\,
    \mathbf Q\cdot\delta\mathbf r ,
\end{equation}
up to an overall sign. Thus, a sliding fluctuation is shifted by one-quarter period relative to the equilibrium CDW profile. The real-space eigenvector of Mode 3 has precisely this structure, identifying it as the sliding mode of the unidirectional CDW. For an incommensurate CDW, this phase translation would be a gapless phason. In the period-8 commensurate approximation, however, the CDW phase is pinned to the lattice, so the mode is weakly gapped. Its anomalously small Hessian eigenvalue therefore indicates exceptionally weak commensurability pinning. And because Mode 3 is mirror odd, a finite static component along this mode would therefore break $M_{x+y}$.